\documentclass[aps,twocolumn,floats,pre,nofootinbib,superscriptaddress]{revtex4-2} 
\usepackage{amsmath, amsfonts}
\usepackage{graphicx}
\usepackage{tikz}
\usepackage{url}
\usepackage{bbm}
 \usepackage{textgreek,bm}

\begin{document}

\title{The FlEye camera: Sampling the joint distribution of natural scenes and motion}

\author{Charles J. Edelson}
\email{Contact Author: cedelson@iu.edu}
\affiliation{Department of Physics, Indiana University, Bloomington, IN 47405}
\author{Paul Smith}
\affiliation{Department of Physics, Indiana University, Bloomington, IN 47405}
\author{Sima Setayeshgar}
\affiliation{Department of Physics, Indiana University, Bloomington, IN 47405}
\author{William Bialek}
\affiliation{Joseph Henry Laboratories of Physics, Princeton University, Princeton, NJ 08544}
\author{Rob R. de Ruyter van Steveninck}
\affiliation{Department of Physics, Indiana University, Bloomington, IN 47405}

\date{\today}

\begin{abstract}
To make efficient use of limited physical resources, the brain must match its coding and computational strategies to the statistical structure of input signals.  An attractive testing ground for these principles is the problem of motion estimation in the fly visual system: we understand the optics of the compound eye, have a quantitative description of input signals and noise from the retina, and can record from output neurons that encode estimates of different velocity components. Furthermore, recent work provides a nearly complete wiring diagram of the intervening circuitry.  What is missing is a characterization of the visual signals and motions that flies encounter in a natural context.  We attack this directly with the development of a specialized camera that matches the high temporal resolution, optical properties, and spectral sensitivity of the fly's eye; inertial motion sensors provide ground truth about rotations and translations through the world.  We describe the design, construction, and performance characteristics of this FlEye camera. To illustrate the opportunities created by this instrument we use data on movies and motion to construct optimal local motion estimators that can be compared with the responses of the fly's motion sensitive neurons.
\end{abstract}

\maketitle

\section{Introduction}

It is an old idea that the algorithms for information processing in the brain should be adapted to the structure of physical signals in the world \cite{helmholtz}.  These ideas became more precise in the mid--twentieth century with the development of information theory \cite{shannon1948mathematical} and theories for signal detection and estimation \cite{wiener_1949,lawson+uhlenbeck_50}.  These approaches are probabilistic, with strong connections to statistical physics \cite{grassberger+nadal_1994}. As Barlow emphasized, this means that neural coding and computation will be driven by the {\em distribution} of signals in the natural sensory environment \cite{barlow_59}.  The resulting ideas of efficient representation and inference continue to have impact on thinking about the brain and about the physics of life more generally \cite{bialek2012,bialek2024}.  The insect visual system provides an important testing ground for these theories \cite{laughlin_1981,hateren_1992,spikesbook,brenner2000adaptive,clark+fitzgerald_2024}.

Despite their small size and ``basic" morphologies, flies and other flying insects are able to perform complex aerial acrobatics with a precision and ease that would make even the most skilled of human pilots jealous. This aerial prowess has led to an interest in understanding how flies both plan and execute their flights. One avenue to answering this question is determining how a fly senses the world around it and then uses its sensory signals to control its many gravity defying feats. This approach has served as a guiding principle for decades of research, revealing a rich collection of sensory signals used in a variety of clever ways to control flight \cite{zbikowski2004sensor, dickson2006integrative, taylor2007sensory, tuthill2016mechanosensation, lehmann2017neural, dickerson2020timing}. The input-output relationships for a number of these systems have been studied in detail, and in some cases this has been explored down to the level of plausible neural control mechanisms for specific flight behaviors \cite{dickson2008integrative, whitehead2022neuromuscular}.

Of its many sensory systems, the fly's visual system may be the most iconic. For the human observer, insect compound eyes are among the most foreign features of their morphology, and indeed the fly visual system operates in a regime quite different from our own.  The optics of the compound eye provide much lower spatial resolution than the lens of the human eye, but fly photoreceptors have much wider temporal bandwidths, giving the insects greater time resolution and allowing faster behavioral responses \cite{de1996light, cao2007linking}. The submillimeter focal length of the fly's optics would be inconvenient for humans, but allows flies to see what is at the end of their legs, useful for animals that walk on their food \cite{stavenga2003angular}.  

The fly visual system also is attractive for its experimental accessibility.   We understand the diffraction--limited optics of the compound eye \cite{barlow_1952,stavenga_2003A,stavenga2003angular,stavenga2004angular}, and extensive quantitative measurements have been made on the signals and noise in photoreceptor cells and their immediate synaptic targets, the large monpoloar cells (LMCs).  Some of the earliest recordings of neural responses to single photons were in insect photoreceptors \cite{lillywhite_1977}, and LMCs act as nearly ideal photon counters up to rates of $\sim 10^6\,{\rm s}^{-1}$ before saturating \cite{ruyter+laughlin_1996A,de1996light}.  Near the output of the fly's visual system, in the lobula plate, is a dense collection of large identified neurons which respond to optical flow patterns and provide motion information that guides flight trajectory planning, pose correction, and spatial mapping \cite{pierantoni1976look, hausen1984lobula, krapp2008estimation, huston2008visuomotor, ullrich2014influence, williams2018blowfly,wei2020diversity}.  Very recent work in fruit flies has mapped the connectivity in all the neurons leading from photoreceptors to the lobula plate, and more \cite{flywire1}.

The responses of motion-sensitive neurons in the lobula plate adapt to changes in the statistical structure of simplified visual inputs, including mean light levels, the dynamic range and correlation times of light intensity variations, the mean and variance of movement velocities \cite{brenner2000adaptive, van2002timing, nemenman2008neural, drews2020dynamic}.  In some cases it has been possible to show that these adaptations serve to maximize information transfer \cite{brenner2000adaptive,fairhall2001}.  Related to this, under some conditions the motion estimates encoded by single cells in the lobula plate reach a precision close to the physical limits imposed by diffraction blur in the compound eye and noise in the photoreceptors \cite{bialek+al1991,ruyter+bialek1995}.  All of this encourages us to think that the physical principles of efficient representation and inference are relevant, at least in this small corner of the fly's brain.  What is missing from this body of work is a compelling characterization of the visual inputs and movements that flies encounter in nature.

Building on earlier efforts, we have made a direct attack on the problem of characterizing the joint distribution of visual inputs and motion statistics by designing and constructing a camera that samples natural scenes in a manner similar to that of a fly's eye \cite{sinha2021optimal}. The FlEye camera, as we have affectionately dubbed it, accomplishes this through a combination of careful optical and electronic design choices, enabling it to  achieve a signal-to-noise ratio orders of magnitude better than the fly photoreceptor, thus giving us nearly the true light intensities that a fly would experience.   The FlEye camera is equipped with an inertial motion unit (IMU) to allow for simultaneous sampling of motion trajectories. Thus, recording a stroll through the woods with the FlEye camera is equivalent to collecting samples from the joint distribution of fly-like visual inputs and motion trajectories. 

With the camera, we will be able to assemble a large database of movies and motion, sampled from a fly eye's view, providing a resource with which to address a number of questions.  In this article we document the design, construction, and calibration of the FlEye camera itself.  We illustrate what can be done with the instrument by revisiting the construction of optimal local motion estimators \cite{sinha2021optimal}.  In doing so, we have uncovered a surprising interaction among nominally orthogonal signals.

\section{Design and Construction}

\begin{figure}
    \begin{minipage}{8.5cm}
    \begin{tabular}{ll}
        (a) & (b) \\
        \includegraphics[width=0.45\textwidth]{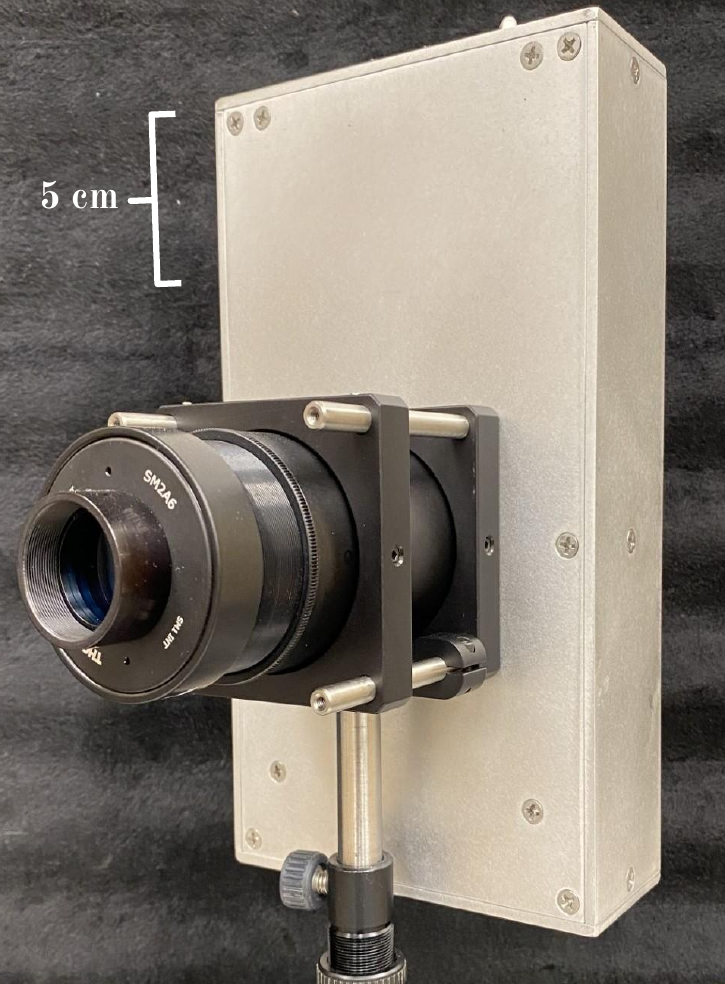} & \includegraphics[width=0.40\textwidth]{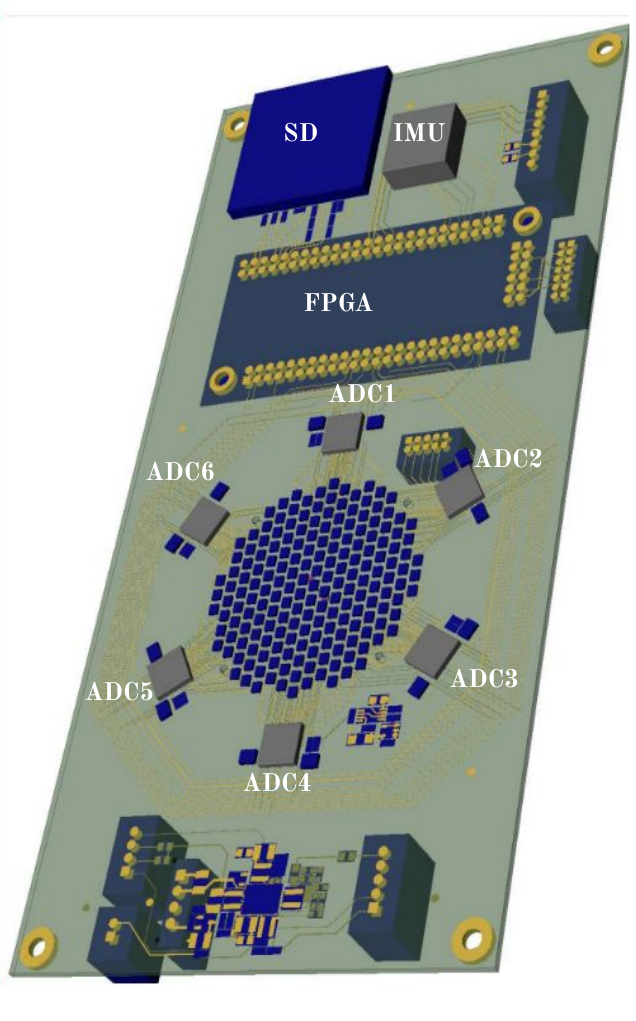} \\
        (c) & (d) \\
        \includegraphics[width=0.45\textwidth]{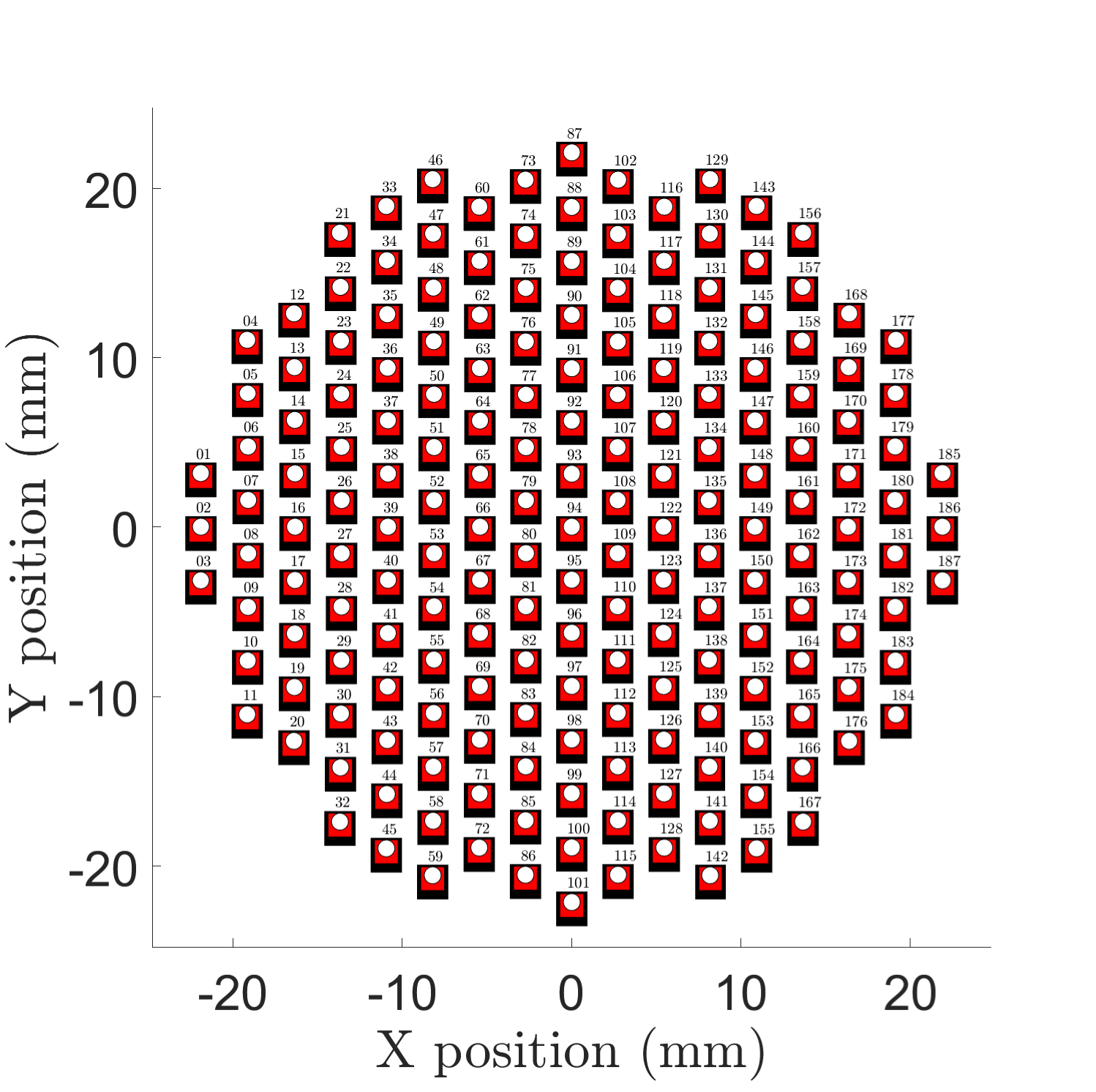} & \includegraphics[width=0.45\textwidth]{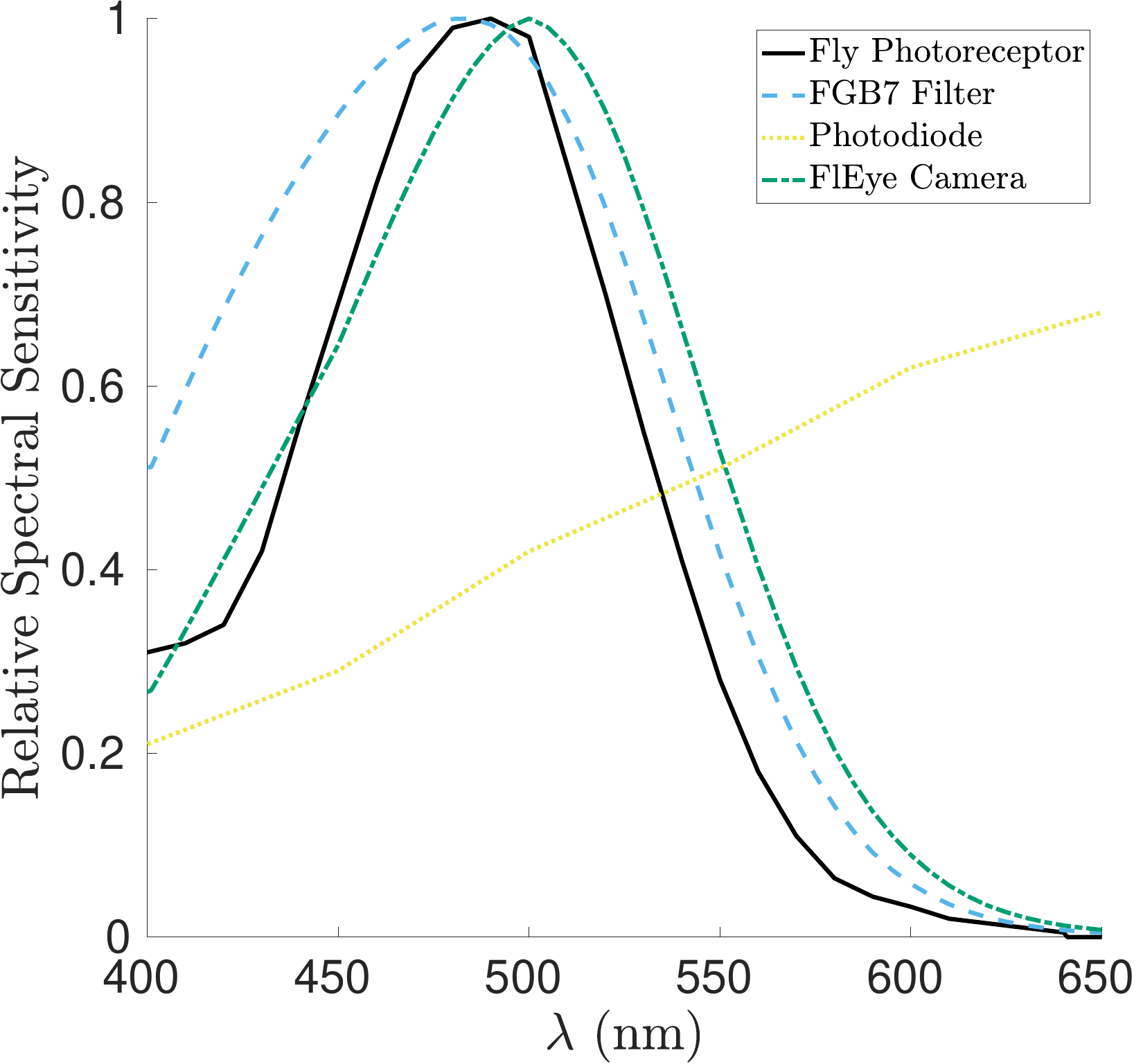}
    \end{tabular}
    \end{minipage}
    \\
    \caption{The FlEye camera. (a) Completed and assembled, its small size is an important design feature, enabling  its use in natural environments. (b) A 3D rendering of the printed circuit board (PCB) with photodiodes and major electronic circuity: analog to digital converters (ADC1-6), field-programmable gate array module (FPGA),  inertial measurement unit (IMU), and Secure Digital card (SD). (c) Photodiode arrangement on the PCB. The photodiodes do not follow a perfect lattice on the xy-plane but instead deviate slightly to compensate for the camera's point projection geometry and ensure constant \textit{angular} separations between the photodiodes. (d) Relative spectral sensitivity of the fly photoreceptor (black), camera photodiodes (yellow), cyan filter (blue), and the FlEye camera (green).}
    \label{fig:fleye_camera}
\end{figure}

The finished FlEye camera is shown in Figure~\ref{fig:fleye_camera}a. For the camera to function successfully  as a sampling device of the joint distribution of natural scenes and motion trajectories it needs to accomplish two major goals. First, the camera needs to record high quality fly-like visual inputs with time-matched motion information. This, at minimum, requires the camera to have a hexagonal photodiode arrangement, a high sample-rate ($>500\,{\rm Hz}$), low electronic noise, and a built-in motion sensor  synchronized with the optical signals. Second, the camera needs to be portable and ergonomic to enable extended recordings ($>30\,{\rm min}$) in a variety of natural environments, such as woods, open fields, and lake shores. Given the uneven footing and variable terrain found in these environments, this requires the camera to be relatively small, light enough for a single experimenter to carry, simple to operate in the field, and to have onboard data storage and power. Thus, the first goal constrains the electronics and optics of the camera while the second goal constrains the body and housing of the camera. 

Working against the goal of compactness is the desire for a large collecting area.  While it makes sense to match the spatial and temporal resolution of the fly's eye, we would like to collect more light and thus have more nearly ground truth about the patterns of light intensity impinging on the retina.  We can always add noise to these data to simulate the signals generated by the fly's photoreceptors.

In addition to these technical constraints, the camera had a more practical constraint: It needs to be affordable to construct and operate. Although commercial high speed cameras exist, they come at high cost and generally with specifications that are unsuitable for our goals. The FlEye camera needed to be relatively inexpensive to construct, have an analysis friendly data format, and accomplish both of these goals while still accurately mimicking the optics of the fly compound eye. 

With these constraints in mind, the camera's design was split into three major components: electronics, optics, and housing, as described in detail below. 

\subsection{Electronics} \label{sec:electronics}

The heart of the FlEye camera is a single printed circuit board (PCB), which is shown in Figure~\ref{fig:fleye_camera}b. The PCB integrates all of the major electronic components of the camera: photodiodes, analog to digital converters (ADCs), inertial measurement unit (IMU), data storage, and power supply. Central to the PCB are 187 OSRAM SFH 2704 silicon PIN photodiodes. The photodiode signals are routed to six Texas Instrument 32-channel current-input DDC232 ADCs (ADC1-6 in Fig~\ref{fig:fleye_camera}b). The ADCs operate at a 1000 Hz sampling rate with an 20-bit output range. An Analog Devices ADIS16500 IMU (IMU in Fig~\ref{fig:fleye_camera}b) records six motion signals: pitch, yaw, and roll angular velocities and the x, y, and z translational accelerations. The IMU uses a 2000 Hz sampling rate, exactly doubling the photodiode sampling rate, and a 16-bit output range. A Trenz Electronics TE0725 field programmable gate array (FPGA) module, equipped with a  Xilinx Artix-7 XC7A35T-2CSG324C FPGA, supplies a common clock for the IMU and ADCs (FPGA in Fig~\ref{fig:fleye_camera}b). Additionally, the Xilinx FPGA serves as the data collation and writing device, collecting all the ADC and IMU signals into 1024 byte blocks before writing them to an external Secure Digital (SD) card using the SimpleSDHC  library \cite{SimpleSDHC}. The FPGA also provides a frame counter and a tag switch (see Section \ref{sec:housing}), which are both included in the 1024 byte write blocks. Finally, power was supplied to the camera with a Jauch LI18650PBF 3.6 volt 6.5 AH lithium-ion battery through a Linear Technology LTC3586-3 high efficiency USB power manager, which also managed charging of the camera battery.

\subsection{Optics and photodiode array} \label{sec:optics}

The camera's optics were designed to mimic the fly's hexagonal sampling raster, point spread function (PSF), and photoreceptor spectral sensitivity. Each photodiode has an active surface measuring $1.4\,{\rm mm} \times 1.4\, {\rm mm}$ and they are arranged into a hexagonal lattice within a 2" circular domain, an arrangement chosen to minimize the number edge photodiodes.  An angular spacing between neighboring photodiodes of $\phi_0 = 1.57\,{\rm deg}$ matches the angular separation of photoreceptors in the blowfly {\em Calliphora vicina} \cite{beersma1977retinal}, and this is achieved by placing the photodiodes on the PCB as shown in Fig~\ref{fig:fleye_camera}c.  We note that this is not  a perfect lattice in the plane of the PCB; slight deviations were necessary to ensure the photodiodes sampled at a constant angular separation, a problem the fly avoids by having hemispherical eyes.  The intrinsic spectral sensitivity of the photodiodes is very broad, but once assembled with other optical elements it matches that of the fly photoreceptors, as shown in Fig~\ref{fig:fleye_camera}d.   Most notably, it has a prominent peak at 490 nm \cite{stavenga2004angular}. Fly photoreceptors have an additional peak in the UV, which for technical reasons we chose to ignore here. The fly PSF approximates a Gaussian with a standard deviation of $\Delta\phi = 0.51\,{\rm deg}$ \cite{smakman+al_1984}, and as described below our optical design matches this quite closely.

\begin{figure}
    \begin{minipage}{8.5cm}
    \begin{tabular}{l}
        (a) \\
        \includegraphics[width=\linewidth]{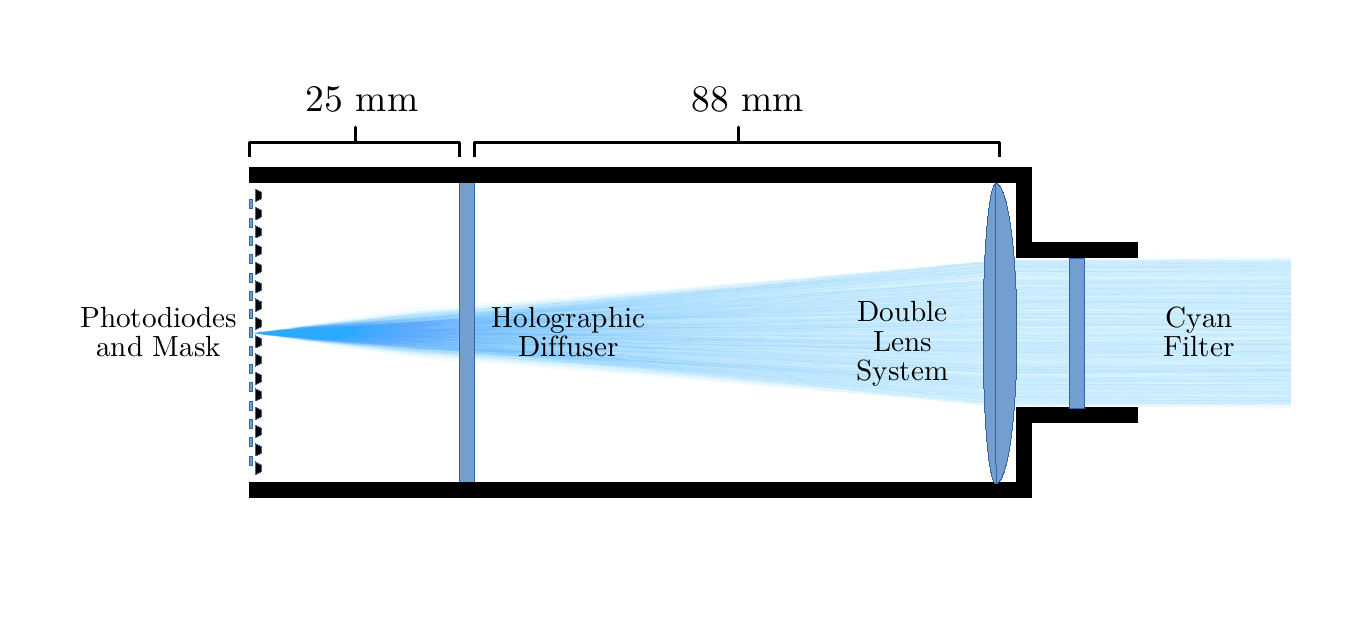} \\
        (b) \\
        \includegraphics[width=\linewidth]{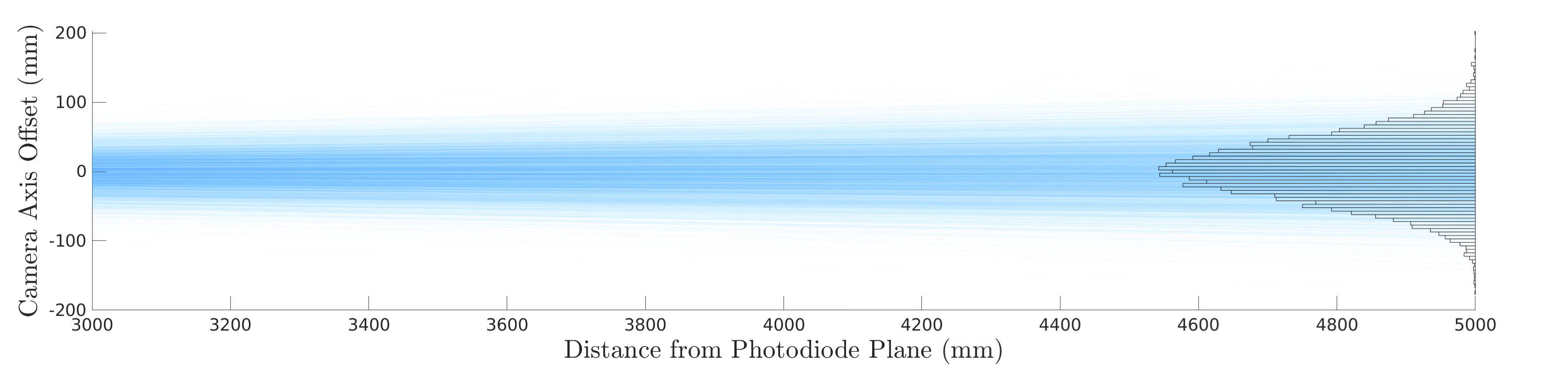} \\
    \end{tabular}
    \end{minipage}
    \centering
    \caption{FlEye camera optics. (a) Schematic  showing the important optical components needed to accurately mimic the fly PSF and spectral sensitivity. (b) Reverse ray tracing simulations of the camera's optics for the central photodiode (photodiode 94). Individual rays are created at the central photodiode before propagating through the camera optics backwards. Only rays that do not intersect with an opaque surface are shown here and ray opacities are set to 1\% to create a ray density plot. The right histogram shows counts of rays incident passing through small regions of the y-axis 5 meters away from the photodiode plane.
    \label{fig:fleye_optics}}
\end{figure}
 
Design and simulation of the optics were done using the MATLAB MCRayTracing library \cite{MCRayTracing}, described in Appendix \ref{sec:calibrations}, before the camera was constructed and calibrated. An overall schematic of the final optical configuration of the camera can be seen in Fig~\ref{fig:fleye_optics}. Starting closest to the PCB, the photodiodes were masked to a 1 mm circular active area to remove any affects their square profile had on the PSF. The mask was milled from a 2" Thorlabs SM2Pl externally threaded plug with an array of $1\,{\rm mm}$ holes drilled to match the lattice of the photodiodes. The mask was mounted directly to the PCB and aligned with the photodiode array by hand. A 2" Thorlabs SM2L internally threaded optical tube was attached to the PCB. Within the tube, a 50 mm Edmund Optics  5 deg full-width-half-max holographic diffuser was placed 25 mm in front of the photodiode plane. The filter was held in position using two Thorlabs SM2RR retaining rings, one on each side of the holographic filter.  

A Thorlabs SM2V15 adjustable lens tube is threaded into the Thorlabs SM2ML tube to facilitate installation and positioning of the camera's main lens system. This came in the form of a 2" diameter $500\,{\rm mm}$ focal length Newport Optics plano convex lens and a $150\,{\rm mm}$ Newport Optics plano convex lens double lens system, creating an effective focal length of $115\, {\rm  mm}$. The SM2V15 was used to position the double lens systems so that their optical center was positioned 113 mm in front of the photodiode plane. This put the photodiode plane slightly out of focus for the lenses, an intended effect. The 2" aperture of the camera was then reduced to 1" using a Thorlabs SM2A6 thread adapter, which was gently snugged up to the double lens system to hold them in place within the SM2V15 tube. A short section of 1" tubing was attached to the thread adapter and a 25 mm diameter Thorlabs FGB7 colored glass cyan bandpass filter ($435 - 500\,{\rm nm}$)  was mounted within this tubing. As with the holographic filter, the cyan filter was held in place using two 1" retaining rings. The combined spectral sensitivity of the photodiode and FGB7 cyan filter is shown in Fig~\ref{fig:fleye_camera}d.

\subsection{Housing} \label{sec:housing}
The final component of the FlEye camera is the housing used to bring everything together. The housing was constructed from aluminum and was cut to comfortably fit the PCB, battery, SD card, a number of control buttons, a charging port, and the optics. Controls of the camera were kept simple, with an SD card slot, power switch, start and stop recording button, and a tag switch. The tag switch allows an experimenter to tag a portion of a recording for later inclusion (or exclusion) in an analysis. Power is supplied by a PSM10A-050 5V power supply and a single green LED light indicates when the camera is fully charged. Finally, the optics are stabilized using a Thorlabs optical railing and cage system constructed from two LCP09 $60\,{\rm mm}$ cage plate and 6" cage assembly rods. These railings have a mounting point for additional Thorlab supports, allowing the camera to be mounted on standard optical railing for calibrations, experiments, and recordings.

\section{Performance Characteristics}
\label{sec:performance}

We made several measurements both for calibration and to be sure that the instrument met our design goals. Specifically, we looked at the photodiode point spread functions (PSFs), the noise power spectrum, the contrast transfer function, and the effective Poisson rate. Additionally, the gryo-optical properties of the camera needed to be measured to ensure proper temporal alignment of motion and optical data. Finally, photodiode response curves needed to be flattened and converted into physical units. Experimental details and mathematical definitions for each of these measurements are included in Appendix \ref{sec:calibrations}.

\begin{figure}[t]
    \includegraphics[width=\linewidth]{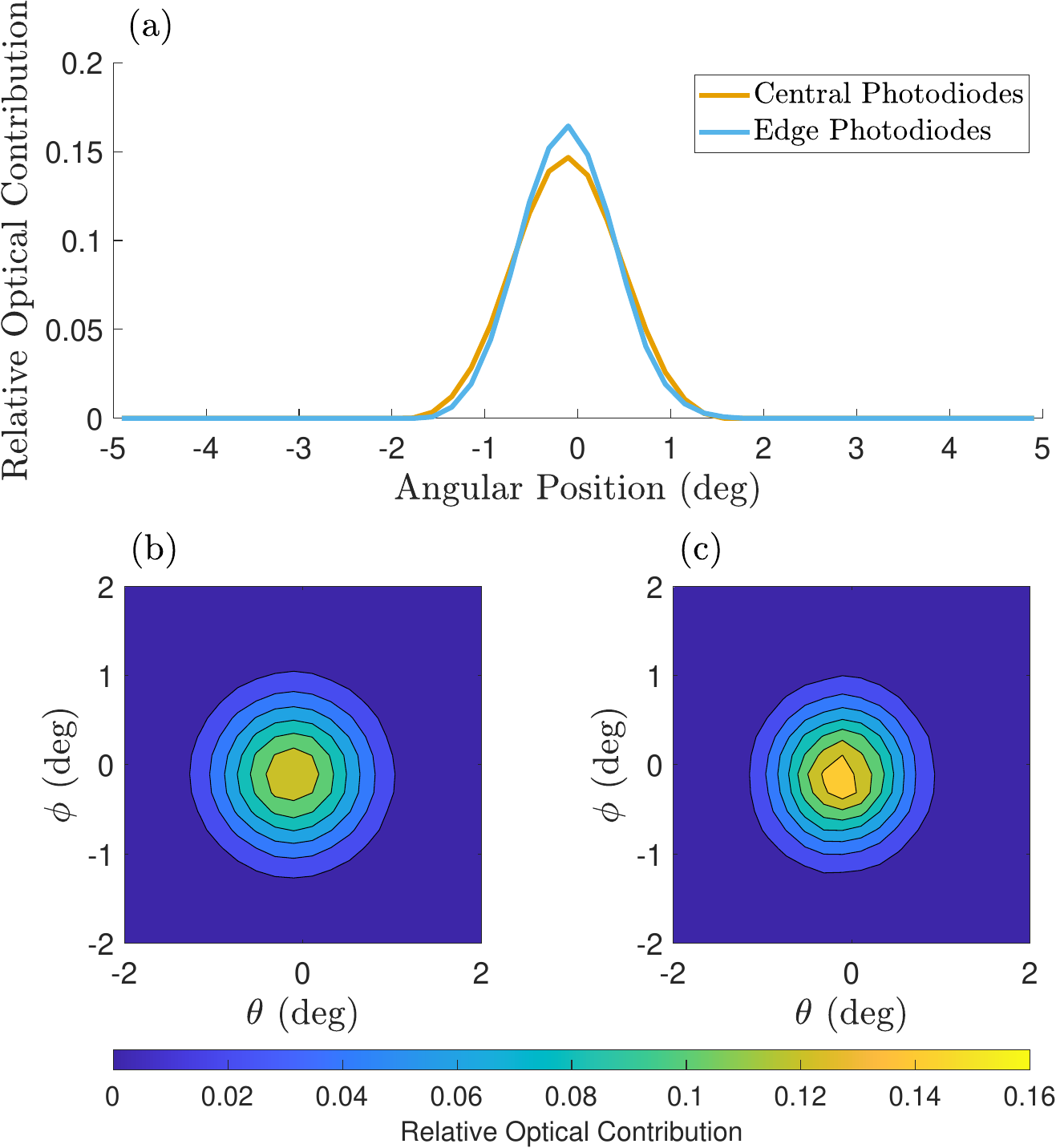}
    \caption{Optically measured average photodiode point spread functions (PSFs) for the FlEye camera. (a) The average central (blue) and edge (orange) PSF marginalized along the $\phi$ angular dimension. (b) Contour plot of the average central photodiode PSF. (c). Contour plot of the average edge photodiode PSF. The measured PSFs in (b) and (c) are both approximately Gaussian. The central photodiodes' PSFs are slightly wider than the edge photodiodes'. Additionally, the edge photodiodes appear slightly more anisotropic, a feature that is reflected in their computed covariance matrices, Eqs (\ref{eq:cov_center}) and (\ref{eq:cov_edge}). Finally, since the PSF are approximately Gaussian, the marginals in (a) also are approximately Gaussian. Notice the slight skew visible in the edge photodiode PSFs compared to the central photodiodes.}
    \label{fig:average_psf}
\end{figure}

\subsection{Point spread functions}

Measurements of the photodiode PSFs for central and edge pixels are shown in Fig~\ref{fig:average_psf}. The average central photodiode PSF (Fig~\ref{fig:average_psf}a, and Fig~\ref{fig:average_psf}b), is Gaussian in profile and very nearly circularly symmetric.  To a very good approximation the PSF thus has the form
\begin{equation}
    PSF(\phi,\theta ) = {1\over Z} \exp\left[ - {1\over 2}
    \left(
    \begin{array}{c}
         \delta\phi\\
         \delta\theta 
    \end{array}
    \right)^T \Sigma_{\rm central}^{-1}
    \left(
    \begin{array}{c}
         \delta\phi\\
         \delta\theta 
    \end{array}
    \right)
    \right]
\end{equation}
where $\delta\phi$ and $\delta\theta$ are angular distances from the central point, in the azimuthal and polar directions, respectively, and $Z$ is a normalization constant.
The measured covariance matrix is
\begin{equation}
    \label{eq:cov_center}
    \Sigma_{\text{central}}  = 
    \begin{pmatrix}
        0.29 \pm 0.0007 & -0.0009 \pm 0.0006 \\
        -0.0009 \pm 0.0006 & 0.30  \pm 0.0009\\
    \end{pmatrix} \, {\rm deg}^2,   
\end{equation}
where the errors are standard errors of the mean. Notice that the diagonal elements are $0.29\,{\rm deg}^2 = (0.54\,{\rm deg})^2$, so that they match the observed width of the fly's PSF. The skewness of the average photodiode PSF, $\gamma_3 = -0.021$, gives a sense for the smallness of deviations from a Gaussian. Meanwhile, the average edge photodiode PSF (blue in Fig~\ref{fig:average_psf}a, and Fig~\ref{fig:average_psf}c), is also Gaussian in profile, with  a covariance matrix of
\begin{equation}
    \label{eq:cov_edge}
    \Sigma_{\rm edge}  = 
    \begin{pmatrix}
        0.26 \pm 0.0029 & -0.003 \pm 0.0006 \\
        -0.003 \pm 0.0006 & 0.28 \pm 0.0111\\
    \end{pmatrix} \, {\rm deg}^2.   
\end{equation}
We see that (1) the widths are slightly smaller than for the central pixels, (2) there is still small but more significant anisotropy measured by the off--diagonal elements, and (3) the skewness is slightly larger at $\gamma_3 = -0.05$.

\begin{figure}[b]
    \centering
    \includegraphics[width=\linewidth]{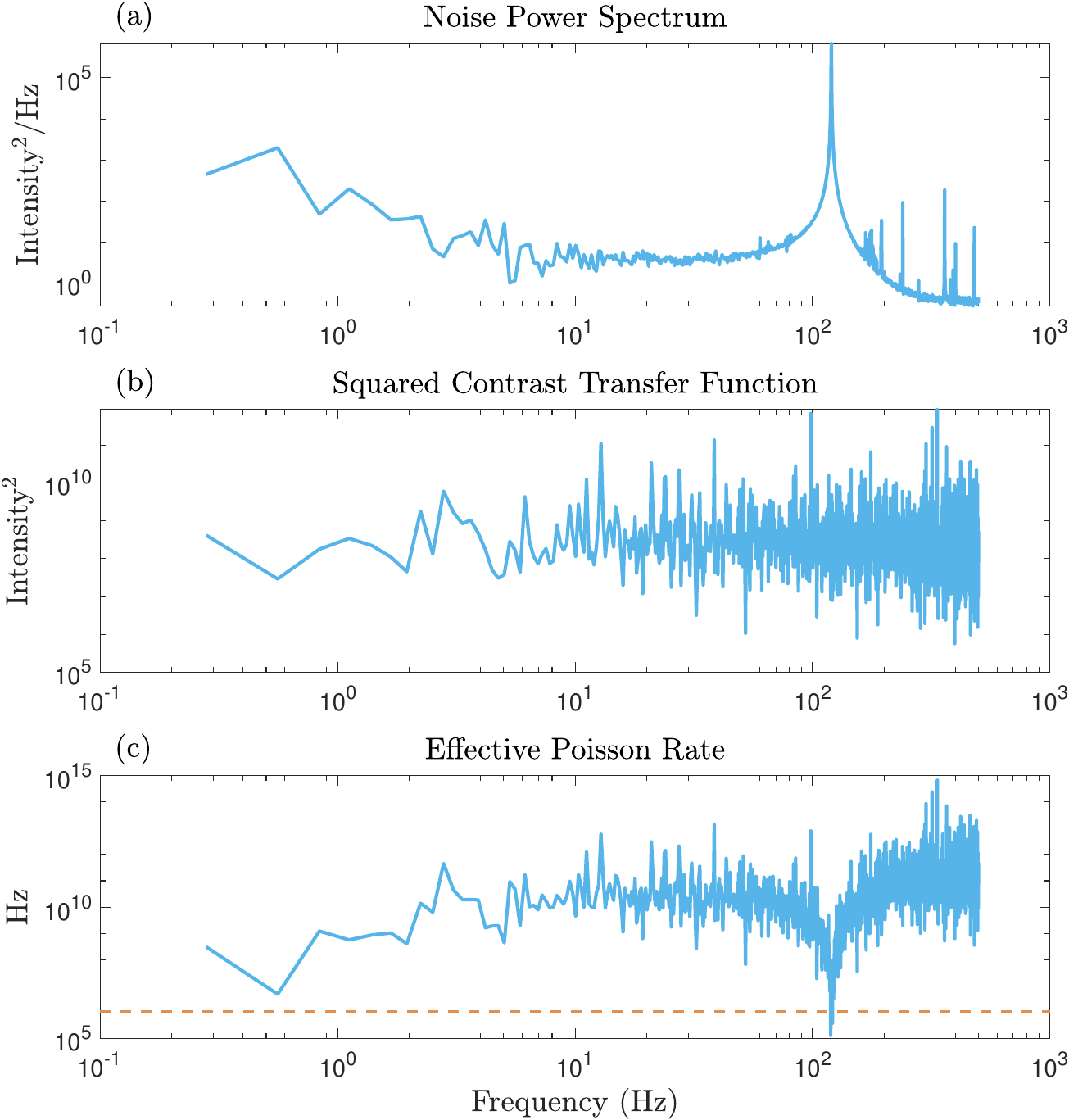}
    \caption{Signal and noise characteristics of the photodiodes. (a) Average noise power spectrum of the photodiode ``intensity'' output, $N_s(\omega )$. (b) Average squared transfer function from contrast to intensity, $|\tilde T (\omega)|^2$.  (c) Average photodiode effective Poisson rate $\lambda_{\rm eff}$ from Eq (\ref{lambda_def}). For comparison, fly photoreceptors and large monopolar cells saturare at $\lambda_{\rm eff} \sim 10^6\,{\rm Hz}$ (orange dashed line).
    \label{fig:nps_transfer_and_lambda}}
\end{figure}

\subsection{Effective noise levels}

The photodiodes of the FlEye camera have an output ``intensity'' $s(t)$ that we have calibrated so that we can report absolute light levels (Appendix \ref{sec:flattening_and_cal}).  The noise in the intensity signal has a power spectrum $N_s(\omega)$ shown in Fig~\ref{fig:nps_transfer_and_lambda}a.  This spectrum is largely flat, rolling off at very high frequencies and having some hints of $1/f$ behavior at the lowest frequencies; the peak at $120\,{\rm Hz}$ is an artefact of having made these measurements with a halogen lamp, which we used in order to calibrate the camera at intensities approaching outdoor levels.  The response of the photodiode intensity $s$ to changes in light intensity $L$ is quite linear, so we can define at each frequency a contrast transfer function $\tilde T (\omega)$ through
\begin{equation}
    \tilde s (\omega ) = \tilde T (\omega ) \tilde L (\omega ) /\bar L ,
\end{equation}
whose squared magnitude is shown in Fig~\ref{fig:nps_transfer_and_lambda}b.  This response is essentially flat across the full frequency range.

If the noise in the photodiode output were generated by random arrival of photons at a rate $\lambda$ then we would have $N_s(\omega) = |\tilde T (\omega )|^2/\lambda$.  Thus we can define the effective Possion rate
\begin{equation}
    \lambda_{\rm eff}(\omega ) = {{|\tilde T (\omega )|^2}\over{N_s(\omega)}} ,
    \label{lambda_def}
\end{equation}
which is shown in Fig~\ref{fig:nps_transfer_and_lambda}c. We see that $\lambda_{\rm eff} \gg 10^6\,{\rm Hz}$ for the camera's entire frequency range (with the exception of the 120 Hz artefact), and $\lambda_{\rm eff} > 10^8\,{\rm Hz}$ for most frequencies.  The relevant comparison is to the effective Poisson rate measured in fly photoreceptors and  LMCs; as we increase light levels these rates saturate at $\lambda_{\rm eff} \sim 10^6 \,{\rm Hz}$  \cite{van2002timing}. We conclude that the FlEye camera provides a view of the world that matches the optics of the fly's eye but at much higher signal-to-noise ratio.

\subsection{Movement signals}

\begin{figure}[t]
    \includegraphics[width=\linewidth]{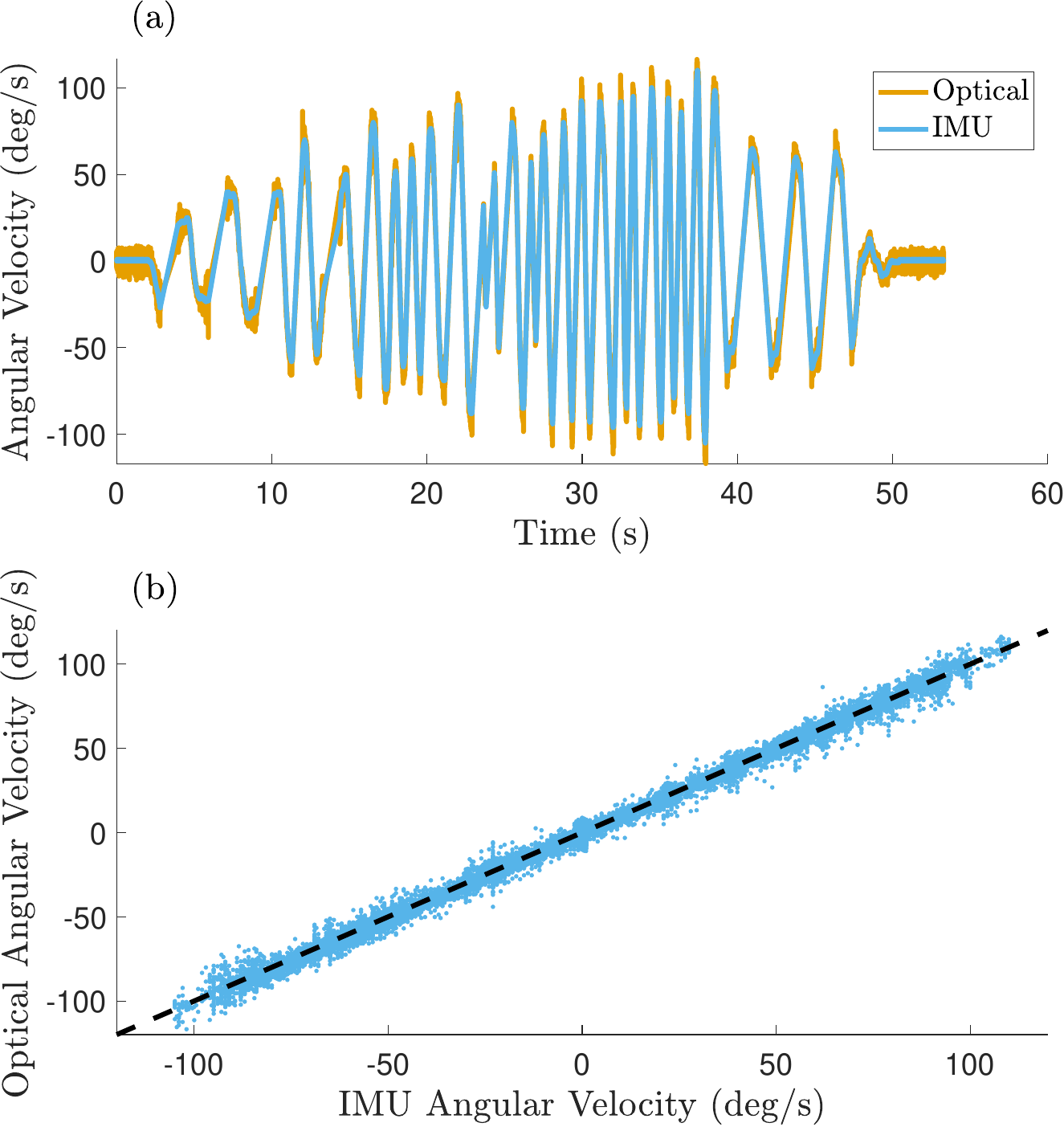}
    \caption{Optical vs direct mechanical measures of yaw velocities. (a) Comparison of optically and IMU measured yaw trace as the camera is rotated in front of a single vertical bar. Optical estimate is derived from the center-of-mass (COM) of light intensity in each frame, and this position vector is differentiated numerically  to compute the COM yaw (orange curve). The IMU measured yaw is the yaw recorded using the FlEye camera's IMU. Notice the two signals align extremely well in time. Additionally, the IMU measured velocity is much smoother than its optical counter part. (b) Scatter plot of IMU and optically measured angular velocity. Black dashed line has a slope of one. The measured data points neatly fall along the dashed line, indicating that the optical and IMU signals are properly in sync and calibrated.
    \label{fig:yaw_com}}
\end{figure}

The inertial motion units (IMUs) provide a direct measure of angular velocity around each of the three axes---pitch, roll, and yaw.  It is essential that these signals be aligned with and calibrated against the optical signals collected by the photodiode array.  As an example of this calibration we expose the camera to a display of a single stationary vertical bar and rotate the camera itself, effectively moving the bar across its field-of-view (FOV). The optical center of mass of each image from this recording was computed and numerically differentiated in time to get an optical measurement of the camera's yaw velocity. This optical measurement is compared to the IMU measured yaw in Fig~\ref{fig:yaw_com}a. The average absolute difference between the two signals is $0.54 \pm 0.03\,{\rm deg/s}$, and the maximum cross-correlation of the two signals is at a lag of zero.

The fact that the maximum cross-correlation occurs at zero lag implies that the FPGA is correctly collating the optical and inertial signals and including them in their correct frames. The difference between the two velocity signals is statistically significant but very small, $\sim 0.25\%$ of the dynamic range in this experiment. The dominant source of this small difference may be noise in the optical estimate, which after all is being made from detectors with a lattice spacing $\phi_0 = 1.57\,{\rm deg}$ and times steps $\Delta t = 0.001\,{\rm s}$.  The reliability of the measurement also is visible in the scatter plot of Fig~\ref{fig:yaw_com}b.

\subsection{Summary}

Taken together, the results of this section indicate that the FlEye camera has met our major design goals.  The spectral sensitivity matches the fly's photoreceptors (Fig~\ref{fig:fleye_camera}d), the optics matches that of the compound eye (Fig~\ref{fig:average_psf}), and the effective signal-to-noise ratio is substantially above the maximum in the fly's retina (Fig~\ref{fig:nps_transfer_and_lambda}).  At the same time the instrument reports angular velocities with millisecond alignment to the image data and better than $\sim 1\%$ precision (Fig~\ref{fig:yaw_com}).

\section{Optimal local motion estimators} \label{sec:ap_I}

The possibility of collecting large calibrated samples of movies and motion, with a fly's eye view, will allow us to address a number of questions about the algorithms for motion estimation.  While our emphasis here is on the FlEye camera itself, we want to give one example of how these data can be used.  To this end we discuss the optimal strategies for estimating motion using local information, which for the fly means comparing the signals from neighboring photoreceptors.  Our results build on earlier work with more limited data \cite{sinha2021optimal}.

\subsection{Theoretical motivation}

If the world moves rigidly at angular velocity $v$ relative to the retina, then the pattern of light intensity will vary in space and time as
\begin{equation}
    I(\phi, t ) = I_0(\phi - vt),
\end{equation}
where for simplicity we consider variations only along the azimuthal angle $\phi$ so that $v$ is yaw velocity.  This suggests that estimating the velocity from visual inputs should be easy, since
\begin{equation}
    v = \hat v_{\rm grad} = - \left[{{\partial I}\over{\partial t}}\right]\left[{{\partial I}\over{\partial\phi}} \right]^{-1}.
    \label{grad1}
\end{equation}
This simple estimator often is called the ``gradient estimator'' \cite{limb1975estimating}.  Naively this is an optimal estimator, solving a mathematically idealized problem.  But in the real world signals are accompanied by noise.  Typically noise extends to higher frequencies than the signal, and thus is amplified by differentiation.  Further, in the presence of noise division is a dangerous operation since fluctuations in the noise could lead to division by zero.  The gradient estimator involves both differentiation and division, and thus can fail dramatically when noise levels are high.

Experiments on the visually guided behaviors of insects are the origin of a very different model for motion estimation, the correlator model \cite{reichardt1961autocorrelation}.  We can write the correlator roughly as
\begin{equation}
    \hat v_{\rm corr} \propto - \left[{{\partial I}\over{\partial t}}\right]\times \left[{{\partial I}\over{\partial\phi}} \right] .
    \label{corr1}
\end{equation}
A crucial prediction of the correlator model, and its cousin the motion energy model \cite{adelson1985spatiotemporal}, is that the scale of variations in light intensity (contrast) is confounded with the velocity of movement.  This systematic error can be seen in fly flight control \cite{reichardt1976visual}, the response of the fly's motion sensitive neurons \cite{egelhaaf1989computational,van1994statistical}, and even in human perception under some conditions \cite{van1984temporal}.  Despite these errors, velocity estimates encoded by the sequences of action potentials in the H1 neuron have a reliability close to the physical limits imposed by diffraction blur in the compound eye and noise in the photoreceptors, and in turn most of the photoreceptor noise is photon shot noise \cite{bialek+al1991,ruyter+bialek1995}.  In this sense the fly makes nearly optimal estimates of visual motion.

The observation of performance near the physical limit motivated a theory of optimal visual motion estimation \cite{potters1994statistical}.  Perhaps surprisingly, the optimal estimator has two regimes:  at high signal--to--noise ratio (SNR) it is approximately the gradient estimator, and at low SNR it is the correlator. This raises the possibility that systematic errors of motion estimation are not a limitation of biological mechanisms but rather the optimal strategy in the face of large noise or ambiguity in the input physical signals. The same idea was used to explain errors in human motion perception \cite{weiss+al2002,stocker+simoncelli2006}, and can be seen as part of a broader view of percepts as Bayesian combinations of noisy sensory data with prior expectations \cite{knill+richards1996}.

It is attractive to think that errors of neural computation are predictable as a consequence of optimal estimation in the presence of physical limitations.  The problem is that quantitative versions of these statements depend on the statistical structure of the incoming signals, and on the effective noise levels in the system.  In the absence of independent access to these quantities we don't really know whether the predicted systematic errors extend into the regime of relevance for experiments on neurons or perception.  This is a problem that we tried to address with an earlier version of the FlEye camera \cite{sinha2021optimal}.  Here we revisit the issue with our new instrument, which collects data at substantially higher efficiency.

\subsection{Data}

We collected data along a forty--five minute walk through the woods of the Porter West Nature Preserve, a Sycamore Land Trust property located near Bloomington, Indiana. The recording was reformatted using the FlEye Reader Python package \cite{FlEyeReader} before being imported into MATLAB, and each photodiode response was calibrated into units of radiance using a two-knot quadratic spline individually fit to each photodiode to minimize the sum of square relative errors; these and other details are described in Appendix \ref{sec:flattening_and_cal}. The 187 optical signals were then moving-average filtered with a window width of 2 frames (2 ms) before being  down-sampled by dropping every other frame. While this reduced the camera's sampling rate to 500 Hz it also improved individual photodiode SNRs.

The gradient [Eq (\ref{grad1})] and correlator [Eq (\ref{corr1})] models make it plausible that velocity estimates that are built from local image data will depend only on the spatial and temporal derivatives of the image.  In practice we estimate the spatial derivative by taking differences between neighboring pixels---as must happen in the real compound eye---and approximate the temporal derivative by the difference between successive frames.  Expecting that the absolute light intensity is irrelevant we use the logarithm of the intensity as the basic signal.  Note that the large dynamic range of the photodiodes is crucial in being able to take logarithms with no corrections. 

\begin{figure}[t]
    \includegraphics[width=\linewidth]{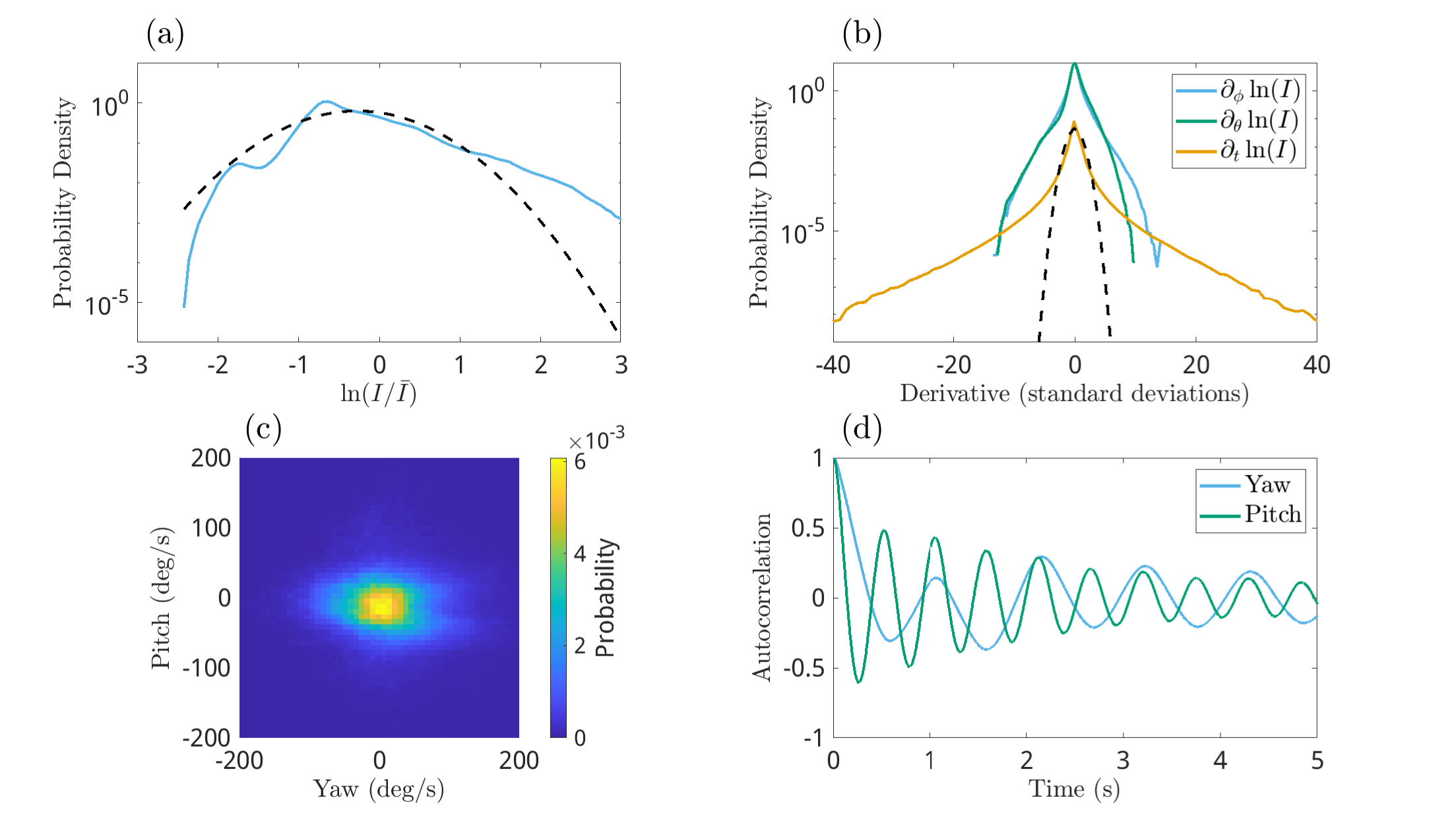}
    \caption{Statistics of the natural scenes recorded on a 45 minute walk through the woods. (a) The distribution of the log intensity contrast signal for the FlEye camera's photodiodes. Dashed black curve is a normal distribution with the same mean and variance. (b) Distributions of the spatial and temporal gradients in units of standard deviations, compared with a standard normal distribution. (c) Distribution of pitch and yaw velocities. (d) Temporal correlation of the velocity waveforms.}
    \label{fig:statistics}
\end{figure}

Figure \ref{fig:statistics} illustrates the basic statistical structure of these local signals.  We see that the distribution of (log) intensity is strongly non--Gaussian, with samples ranging over nearly three decades (Fig~\ref{fig:statistics}a).  The distributions of derivatives have even longer, nearly exponential tails (Fig~\ref{fig:statistics}b), with the time derivatives, in particular, spanning $\pm20$ standard deviations.  While human walking does not reach the enthusiasm of fly flight, angular velocities are on the scale of $\sim 100\,{\rm deg}/{\rm s}$, again with long tails to the distribution (Fig~\ref{fig:statistics}c).  On this particular walk the angular velocities have an oscillation on the $\sim 1\,{\rm s}$ time scale, with pitch and yaw periods differing by a factor two (Fig~\ref{fig:statistics}d).  While this corresponds to relatively gentle movements, we note that the autocorrelation function of the velocities is sharp at zero lag, implying that local accelerations are very large and the waveforms are rough on short time scales.
 
\subsection{Constructing the estimator}

In general if we are trying to estimate a feature $f$ based on data $X$, we should consider the broadest possible class of estimators $\hat f(X)$. But if we want the estimator that makes the smallest mean--square error then we should compute the conditional mean \cite{bialek2012}:
\begin{eqnarray}
\min_{\hat f} \langle | \hat f(X) - f|^2\rangle \Rightarrow \hat f(X) &=& \int df\,f P(f|X)\\
&=& \mathbb{E}[f | X] .
\end{eqnarray}
In the present case the feature of interest is the angular (yaw) velocity, $v_\phi$, and the relevant data are the local temporal and spatial   derivatives, $\partial_t \ln I $ and $\partial_\phi \ln I$.   The optimal estimator then is
\begin{equation}
    \hat{v}_\phi = \mathbb{E}[v_\phi| \partial_t \ln I, \partial_\phi \ln I] 
    \label{optlocal1}
\end{equation}
The key idea is to replace the average over the distribution $P(v_\phi| \partial_t \ln I, \partial_\phi \ln I)$ with average over samples collected by our instrument, in the spirit of Monte Carlo integration \cite{sinha2021optimal}.  To compute conditional means we place the estimated derivatives into $N=100$ equal sized bins along each axis, and the smoothness of our results suggests that there is no significant structure hiding within the bins.

\begin{figure}[t]
    \includegraphics[width=\linewidth]{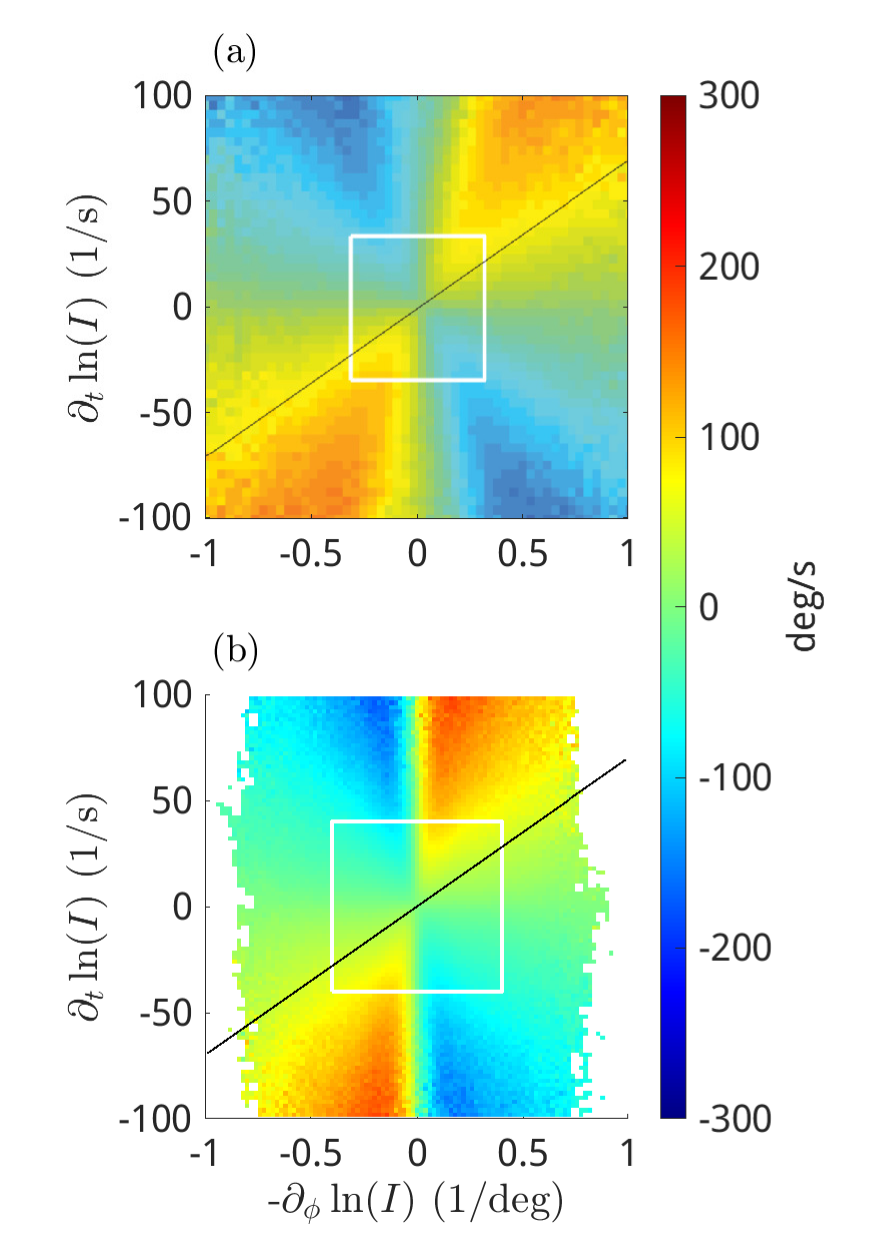}
    \caption{Comparison of the optimal estimators of (yaw) angular velocity. (a) The estimator computed in \cite{sinha2021optimal} using a dataset collected with the old camera. The white box encloses 90\% of the data and the black line is the $\hat{v}_{\rm grad} = 70$ deg/s. Notice that at low values of $\partial_t\ln(I)$ the estimator appears correlator-like. Meanwhile at higher values of the temporal derivative the estimator is gradient like. (b) The estimator computed using the new FlEye camera dataset recorded on the 45 min walk through the woods. As before, the white box encloses 90\% of the data and the black line is the $\hat{v}_{\rm grad} = 70$ deg/s. The two estimators are qualitatively similar, with the same correlator to gradient transition happening as derivatives become larger.
    \label{fig:mmse_yaw_estimator}}
\end{figure}

Results for the optimal estimator defined in Eq (\ref{optlocal1}) are shown in Fig \ref{fig:mmse_yaw_estimator}, where we also compare with previous results \cite{sinha2021optimal}.  We see that where derivatives are large the contours of constant $\hat{v}_\phi$ are straight lines through the origin of the $(\partial_t \ln I, \partial_\phi \ln I)$ plane, as expected from the gradient model in Eq (\ref{grad1}).  As derivatives become smaller the contours of constant estimated velocity bend into hyperbolae, as expected from the correlator model in Eq (\ref{corr1}).  This crossover occurs at modest values of the spatial gradient, and thus is significant on the scale of typical signals encountered on a walk through the woods.   This basic pattern was visible in the earlier data (Fig \ref{fig:mmse_yaw_estimator}a) but is much clearer in the data with our new instrument (Fig \ref{fig:mmse_yaw_estimator}b).  Part of the difference is that we now just have more data, but it also is important that the larger dynamic range of the photodiodes allows for better sampling in the tails of the intensity distribution, and hence cleaner results across the full plane of derivatives.

We emphasize that the crossover from gradient--like to correlator--like behavior  is visible in the optimal estimator despite the fact that the representation of the light intensity itself is effectively noiseless on the scale of the fly photoreceptors.  We have constructed the optimal estimator from signals collected in a camera with an effective photon counting rate of more than $10^8\,{\rm s}^{-1}$, while the effective counting rates in photoreceptors and their postsynaptic targets saturates at $\sim 10^6\,{\rm s}^{-1}$ \cite{de1996light,ruyter+laughlin_1996A}.  If we add noise to the measured light intensities then the region of correlator--like behavior expands, which is qualitatively consistent with experiments on the responses of motion--sensitive neurons \cite{deruyter1996}.  We will return to a quantitative exploration of this effect in subsequent work.

\begin{figure}[t]
    \centering
    \includegraphics[width=\linewidth]{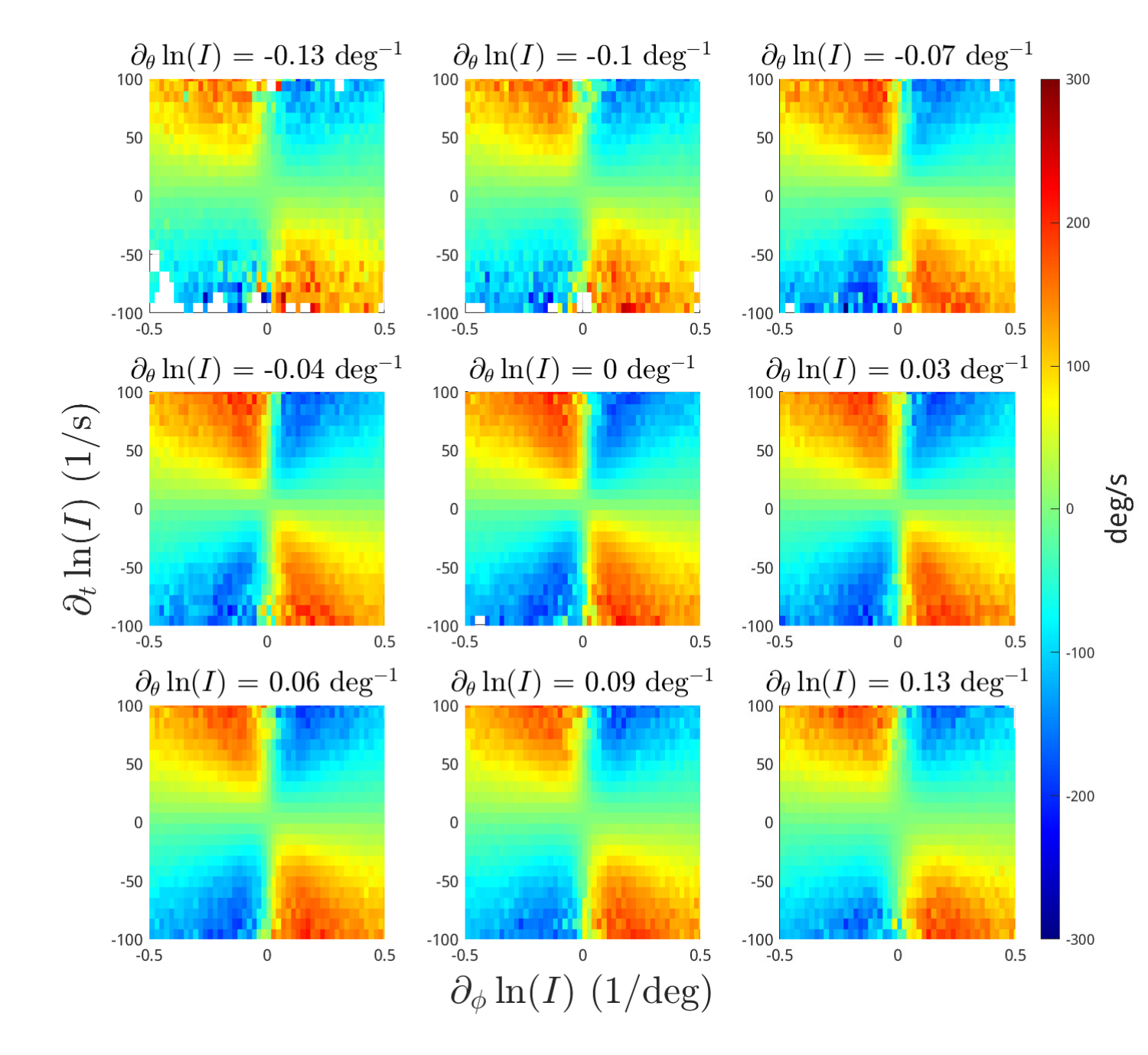}
    \caption{The optimal estimator of yaw velocity from Eq (\ref{eq:MMSE_simul_yaw}). The estimator $\hat v_\phi$ is a scalar field within the volume of $\partial_t \ln(I) \times \partial_\phi \ln(I) \times \partial_\theta \ln(I)$ space. Each panel shows a slice of this estimator at constant  $\partial_\theta \ln(I)$, with values increasing from the upper left. For low values of $|\partial_\theta \ln(I)|$ the  estimator looks similar to the yaw only estimator (Fig \ref{fig:mmse_yaw_estimator}), while at high values of $|\partial_\theta \ln(I)|$ the structure becomes less defined.}
    \label{fig:mmse_simultanious_yaw}
\end{figure}

With better sampling we can afford to divide our data more finely, revealing additional dependencies. The convective or material derivative of light intensity depends on the velocity vector $\vec{v}$ of the camera's motion as
\begin{equation} 
    \frac{dI}{dt} = \frac{\partial I}{\partial t} + \vec{v} \cdot \vec{\nabla} I. 
\end{equation} 
Therefore, we expect the estimate of a given velocity component to also depend on spatial variation of the intensity in the other orthogonal directions. To probe for this effect we consider a generalization of Eq (\ref{optlocal1}) to

\begin{equation}
    \label{eq:MMSE_simul_yaw}
    \hat{v}_\phi = \mathbb{E} [v_\phi| \partial_t \ln I, \partial_\phi \ln I, \partial_\theta \ln I] .
\end{equation}
Figure \ref{fig:mmse_simultanious_yaw} show slices through this function at fixed values of the polar derivative. 

For low values of $\partial_\theta \ln I$ the optimal estimator depends on $\partial_t\ln I$ and $\partial_\phi\ln I$ and looks very much like the ``yaw only'' estimator in Fig \ref{fig:mmse_yaw_estimator}, though perhaps a bit more sharply defined.   As the absolute value of $\partial_\theta \ln I$ increases the contours of constant estimated velocity  move farther from the origin and the ill--defined region at the center of the plane expands.   This implies that the three dimensional contour of the simultaneous yaw estimator is forming a cone within the gradient volume. In addition to the cone--shaped contours, the estimator appears to be ``fuzzier" for higher absolute values of $\partial_\theta \ln(I)$. This suggests that large gradients in the polar or pitch direction, in combination with pitch movement, act as noise for estimating yaw velocity, and we see hints of this paradoxical behavior in neural responses \cite{roy2015encoding}.  A corollary is that motion estimates should be influenced by orthogonal derivatives, and this should be traceable in the patterns of connections to the motion-sensitive neurons \cite{flywire1}.

\section{Conclusion}

Quantitative experiments on visual estimation of motion date back to the early years of the twentieth century \cite{wertheimer_1912}. The mathematical description of the underlying algorithms received an important stimulus from midcentury experiments on insect behavior \cite{hassenstein+reichardt_1956}, and this continues to be a productive example for the exploration of neural computation \cite{clark+fitzgerald_2024}.  Although there is considerable interest in the adaptation of these computations to the structure of natural inputs, what has been missing from this discussion is a body of calibrated data on movies and movements under natural conditions.

In this work, we have reported on the development of the FlEye camera which accurately replicates many of the important features of the fly early visual system while simultaneously capturing high quality motion information. Furthermore, the design of the camera features ease of construction, calibration, and operation.  Thus, the camera is uniquely suited to address questions related to the estimation of velocity from visual information, a problem the fly continuously solves while navigating a complex and dynamic world and which provides an example for neural computation more generally.

To illustrate the potential of FlEye camera we revisited the problem of optimal motion estimation based on local image derivatives in space and time.  We reproduced the predicted crossover from correlator--like to gradient--like estimation, emphasizing that this happens on scales relevant to real world vision.  Our new larger data set allowed us to uncover the contribution of orthogonal signals to the estimation of velocity, e.g. the contribution of pitch gradients to yaw estimation.  This predicts that local motion sensitive neurons that are sensitive to yaw should be receiving inputs from receptors that are separated along the pitch axis, which might otherwise seem irrelevant.

Taken together, these results emphasize that the physical limits to visual estimation are relevant not just for photon counting on a dark night but also in the noonday sun.  The possibility of collecting even larger data sets, under more varied conditions, sets an agenda for more systematic exploration of how optimal local estimators should be combined into wide-field estimators, and how these algorithms should adapt to more subtle statistical features of sensory input.

\begin{acknowledgments}
This research was supported in part by Lilly Endowment, Inc., through the Indiana University Pervasive Technology Institute, and by Indiana University. Additional support was provided by the National Science Foundation, through the Center for the Physics of Biological Function (PHY--1734030), and by fellowships from the Simons Foundation and the John Simon Guggenheim Memorial Foundation (WB). 
\end{acknowledgments}

\appendix

\section{Simulations of FlEye optics}

The MCRayTracing library is a custom built lightweight Monte Carlo ray tracing library designed to simulate two-dimensional ray optics for a number of common optical components \cite{MCRayTracing}. The probabilistic nature of the Monte Carlo ray tracer allows it to easily model diffuser optics with a slight added cost in simulation time. A variety of potential optical set-ups were tested using  reverse ray--tracing experiments before a final camera design was settled upon. These experiments began by instantiating two sets of objects: optical components and light sources. Optical components included lenses, diffusers, opaque panels, and screens while light sources are uniform random emitters of a single ray within a specified angular window. For reverse ray-tracing experiments, light sources were instantiated where the camera photodiodes would be and light rays propagated backwards through the camera optics before finally arriving at a recording screen. Optical objects were instantiated with specific propagation rules for incident rays, allowing them to mimic physical optical systems. Reverse ray-tracing experiments were run for 10,000 iterations with each iteration having a maximum ray propagation depth of 4 objects. Reverse ray--tracing experiments were run to test different lens positions (113 mm vs 115 mm), diffuser FWHM  angles ($5 - 25\,{\rm deg}$), camera apertures (1" and 2"), and PSF convergence for optimal recording distances. Additional simulations were run to determine the expected difference between central and edge photodiode PSFs.

\section{Data Processing}

\begin{figure}[b]
    \includegraphics[width=\linewidth]{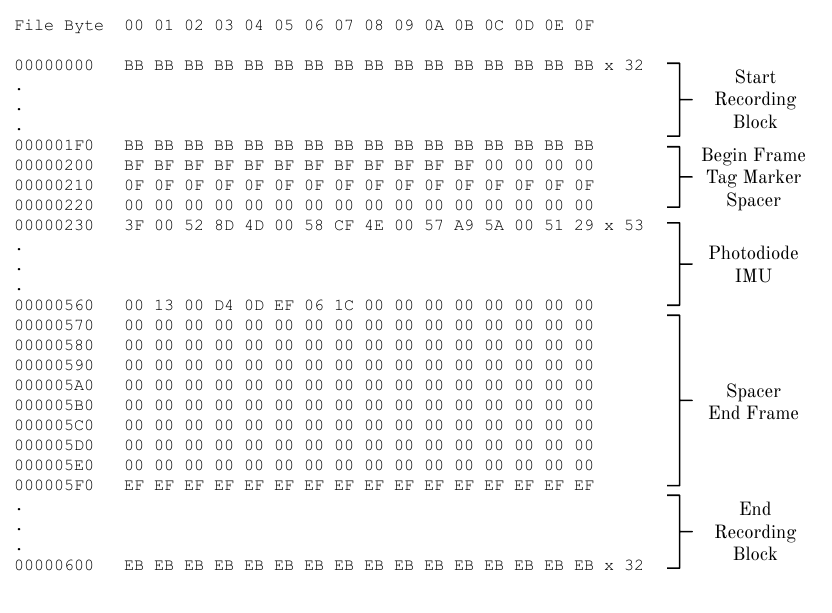}
    \caption{Raw write data format for the FlEye camera. Individual frames are written in 1024 byte chunks, which corresponds to two SD card 512 byte write blocks.
    \label{fig:fleye_data_format}}
\end{figure}

FlEye camera data were written directly to a SanDisk Extreme Secure Digital (SD) card using the FPGA, as discussed in Section \ref{sec:electronics}, and were formatted as shown in Fig~\ref{fig:fleye_data_format}. Camera data were padded to fill out a 1024 byte block to take advantage of the standard SD card write block size of 512 bytes. The SanDisk's Video Speed Class 30 was necessary to facilitate the fast write speeds of the FlEye camera.

While the data format shown in Fig~\ref{fig:fleye_data_format} minimizes the number of write blocks necessary, its mixing of integer types and overall structure makes it difficult to read using standard MATLAB file I/O methods. To overcome this difficulty, the FlEye Reader Python package was developed to read data from raw FlEye recordings \cite{FlEyeReader}. The package contains Python classes for reading, validating, and writing raw FlEye camera SD card data. The data are reformatted into 32 byte integers and written into contiguous frames. Logging is built into the classes and used to note any errors in FlEye camera raw data formats. In addition to these classes, the package contains a simple command-line tool that makes reformatting FlEye recordings into MATLAB friendly formats extremely simple. Finally, the CameraDataLoader MATLAB class was developed to facilitate repeatable loading of reformatted FlEye camera data files.

\section{Performance measurements}
\label{sec:calibrations}

Here we provide some of the technical details behind the meausurements reported in Section \ref{sec:performance}.

\subsection{Optical Measurements}

The photodiode PSF was measured by recording the camera's responses to a $48 \times 48$ high intensity LED display with a $500\,{\rm Hz}$ rate and 12-bit pulse width modulation (PWM). Individual display LEDs had a square profile with an edge length of $5\,{\rm mm}$, giving the full display a height and width of $24.2\,{\rm cm}$. The photodiode PSF and effective focal length were measured simultaneously. 

The camera was placed $138\,{\rm cm}$ from the display, giving each LED an angular dimension of $0.2\,{\rm deg}$, and was oriented so that the camera's field-of-view (FOV) was centered on the LED display. Given that the angular FOV of the camera is $22.8\,{\rm deg}$, the display only filled the central portion of the camera's FOV. This set-up ensured that each camera photodiode's PSF encompassed multiple display LEDs. After the camera was positioned, the lab was darkened and a FlEye recording was started on a fresh SD card. The screen played 100 maximum intensity frames (200 ms) before individually illuminating each display LED at a PWM intensity of 255 for 20 frames (40 ms). LEDs were illuminated starting in the upper left corner of the display and the lighted LEDs moved row-wise before column-wise, ending with the bottom right display LED. The stimulus video ended with another 100 maximum intensity display frames, effectively sandwiching the moving pixel stimulus between two sets of calibration frames. These high intensity frames then served as reference times in the FlEye camera recording and allowed for synchronization between the camera recording frames and the display stimulus times. This procedure was repeated with the camera turned so that the center of the display was in the FOV of the edge photodiodes.

We write the output intensity of the $i^{\rm th}$ photodiode in the FlEye camera at frame $t$ as $s_i(t)$, and the intensity of the $j^{\rm th}$ display LED at frame $t$ as $w_j(t)$.  Then the PSF for the $i^{\rm th}$ photodiode, as a function of screen position $j$, can be estimated as
\begin{equation}
\label{PSF}
    \hat{PSF}_{i,j} = {1\over{Z_i}}\text{Corr}_t [w_j(t), s_i(t)] 
\end{equation}
where Corr$_t[\cdot, \cdot]$ denotes the equal time correlation between the two arguments. The normalization $Z_i$   is given by
\begin{equation}
    Z_i = \sum_j \hat{PSF}_{i,j},
\end{equation}
ensuring the PSF has unit mass. These PSFs can be turned into functions of space, rather than LED, by noting that each screen LED has an associated angular position, $\vec{x}_j$. Thus the value at an arbitrary spatial point within this lattice can be assigned by interpolation. 

Because of the normalization we can treat the PSF as a probability distribution for angular positions.  Thus for each photodiode we can compute a mean position and a covariance matrix, as reported in Eqs (\ref{eq:cov_center}) and (\ref{eq:cov_edge}).  In addition we can estimate the  skewness or dimensionless third cumulant, which provides a measure of the asymmetry of the PSF to its width; this was used to ensure the PSF's were not unexpectedly heavy tailed or asymmetric.

\subsection{Signals and noise}

One of the major design goals of the camera was for it to have a SNR significantly higher than the fly early visual system. In particular, we wanted the camera's SNR to exceed that of the fly large monopolar cell (LMC), a secondary visual neuron that combines the responses of six photoreceptors. We quantify the camera's SNR by looking at the effective Poisson rate, $\lambda_{\rm eff}(\omega)$, as defined in Eq. \ref{lambda_def}. To measure this, we first needed to measure the camera's noise power spectrum $N_s(\omega )$ and contrast transfer function $\tilde T (\omega)$. This was accomplished using two experiments. 

In the first experiment, the camera was placed directly in front of a high-intensity halogen lamp with a light diffuser. The camera was positioned so the halogen lamp filled the entire FOV. This ensured that the lamp served as an extended light source and that each photodiode sees the same radiance. The room was then darkened and a short ($\sim 4\,{\rm s}$) camera recording was captured on a fresh SD card. The NPS was then computed from this data set. We separate the time dependent signal $s_i(t)$ from each photodiode into mean and noise,
\begin{equation}
    s_i(t) = \bar{s}(t) + e_i(t),
\end{equation}
where the mean at each time is computed across the entire array of $N_{\rm pd}$ photodiodes,
\begin{equation}
    \bar{s}(t) = \frac{1}{N_{\rm pd}}\sum_{i = 1}^{N_{\rm pd}} s_i(t) .
\end{equation}
The noise power spectrum $N_s(\omega)$ then was estimated as an average of the squared modulus of the discrete Fourier transforms of the noise waveforms $e_i(t)$:
\begin{equation}
    {N}_s(\omega_n) = \frac{1}{N_{\rm pd}} \sum_{i=1}^{N_{\rm pd}} |\tilde{e}_{i}(\omega_n)|^2.
\end{equation}

In the second experiment, first the FlEye camera's optics were removed, exposing the photodiode array. The camera was then placed $56.9\,{\rm cm}$ in front of a high-intensity LED driven by a $5\,{\rm kHz}$ current source, ensuring the LED was centered within the camera's FOV. A new recording was then started on an empty SD card. The LED display a 1 second constant intensity bright flash before playing 30 repeats of a 5 second spectrally flattened white noise signal. Following the 30 repeats (150 s) the LED again displayed a bright 1 second constant flash. These bright flashes served as calibrations points for comparing the FlEye recording with the known stimulus signal.

The FlEye camera recording was matched to the stimulus timing and reshaped into individual 5 second repeats using the calibration frames. Thus, for the $i^{\rm th}$ photodiode, its signals now takes the form of a 30 x 5000 matrix: $S_i(m; t)$, the signal of photodiode $i$ at time $t$ in the $m^{\rm th}$ stimulus repeat. The signal was standardized across photodiodes to account for differences in individual photodiode gain. 

The contrast transfer function, $\tilde T_i(\omega )$, is defined as the Fourier transform of the photodiode impulse response to a contrast signal, and was estimated for a single photodiode as
\begin{equation}
    \tilde {T}_i(\omega_n) = \frac{1}{30} \sum_{m=1}^{30} \frac{\tilde{S}_i (m; \omega_n)}{\tilde{w}_i(m; \omega_n)},
    \label{Tdef}
\end{equation}
where $\tilde{w}_i(m; \omega_n)$ is the discrete Fourier transform of the contrast stimulus on the $m^{\rm th}$ repeat, in units of the mean LED intensity, and we allow the possibility that different photodiodes see different stimuli.  However, accurately computing this stimulus at high frequencies was difficult due to the lack of a shared clock between the camera and stimulus LED, which introduces high frequency jitter into the transfer function. This was overcome by instead using an estimated stimulus signal, based on averaging the signals of all but the $i^{\rm th}$ photodiode:
\begin{equation} \label{stimuls_estimate}
    \hat{w}_{i}(m;\omega_n)  = \frac{1}{N_{pd} - 1}\sum_{j \neq i}\ \tilde S_j(m;\omega_n ).
\end{equation}
We then replace $\tilde w \rightarrow \hat w$ in Eq (\ref{Tdef}).  Finally we esitmate the effective Poisson rate, averaged over photodiodes: 
\begin{equation}
    \lambda_{\rm eff}(\omega_n) = \left(\frac{\bar{s}_\text{N}}{\bar{s}_\text{T}}\right)^2
    {1\over{N_{\rm pd}}}\sum_i \frac{|\tilde {T}_i(\omega_n)|^2}{N_s(\omega_n)},
\end{equation} 
where the prefactor corrects for the difference in light intensity between the two experiments.

\subsection{Measurements of motion}

The gyro-optical properties of the camera were measured in two experiments. First, a simple rotation test of the FlEye camera on a high precision turntable. The FlEye camera was centered on the table along one of its main rotational axes. The table was then rotated through a half rotation. The IMU's response was integrated along the appropriate rotation axis to test its angular accuracy. 

For the second experiment, the camera was held in front of a high intensity vertical bar light source, approximately one meter away. The room was darkened and a new FlEye recording was started on a fresh SD card. The camera was then gently oscillated from side-to-side by hand for a few seconds before ending the recording. Following this, the optical center of mass of the camera (COM) was computed as
\begin{equation}
    \vec{x}_{\rm COM}(t) = \frac{1}{Z(t)} \sum_{i} s_i(t)\vec{x}_i
\end{equation}
with $Z(t) = \sum_i s_i(t)$ and the vectors $\vec{x}_i$ are the coordinates of the $i^{\rm th}$ photodiode in angular space. This was then numerically differentiated to compute an optically derived estimate of the camera's yaw, $v_{\rm COM}$. 
In Fig~\ref{fig:yaw_com} this optical estimate is compared with the direct readout of the IMU.  

\section{Photodiode flattening and calibrations}
\label{sec:flattening_and_cal}

Before the FlEye camera could be used for experiments the 20-bit photodiode outputs need to be calibrated into physical units and flattened. Flattening  requires all photodiodes to have the same output value for a given local input light intensity. In other words, if the camera viewed a constant intensity homogeneous scene all photodiodes should return the same output value if properly flattened. Photodiode flattening and physical unit calibration were both achieved with a single series of experiments using an extended source. 

A high-intensity halogen lamp with a light diffuser was used as the extended source for measuring photodiode response to fixed homogeneous scene intensities. The lamp's intensity was controlled using a variac and its radiance was measured using a photometer in units of ${\rm mW}/{\rm cm}^2/{\rm sr}$. A WBS 480  cyan filter was used with the photometer to match its spectral sensitivity to that of the FlEye camera. The camera was placed $29\,{\rm cm}$ in front of the extended source with the FOV centered on the halogen lamp. The room was darkened and a new FlEye recording was started on a fresh SD card. Starting with the variac at $100\%$, the radiance of the extended source was measured before starting a FlEye recording. After a few seconds the recording was stopped. The extended source's intensity was decreased by decreasing the variac's setting by 5\% and the process was repeated. Once all the variac intensities had been stepped through, a neutral density 1 (ND1) filter was placed in front of the FlEye camera and photometer and the process was repeated. This was again repeated for an ND2 filter, resulting in radiance measurements spanning over two orders of magnitude. There was now a single FlEye recording associated with each extended source radiance value and each of these recordings consisted of a short time trace of measured photodiode intensities for each FlEye photodiode. Each of these time traces was reduced to a single average measured photodiode intensity to create a final data set of extended source radiances and associated individual photodiode measured intensities. As a note, the light intensity from the halogen lamp exhibited a $120\,{\rm Hz}$ ripple, and care was taken  to ensure that our averaging windows always contained an integer number of these cycles.

With radiance response curves in hand, the problem of flattening and calibrating the photodiode signals reduced to a model selection problem. This was tackled in two parts:  first, selecting an overall model and second, selecting an individual photodiode fitting method. Seven potential models were considered, with linear regression serving as a baseline. Their properties are summarized in Table \ref{tab:models}. Since photodiodes typically have two response regimes, a low-light non-linear and a high-light linear response regime, a variety of splines were investigated as potential fit models given their natural ability to split their input domain into different response profiles. Model complexity was adjusted both by adding knots and by increasing maximum model order. The most complex model considered was a two-knot cubic spline (Cubic 2 Spline in Table \ref{tab:models}).  All models were fit with ordinary least squares (OLS) regression and the best model was selected based on its Akaike information criterion (AIC) distribution across photodiodes. Knot position was set by hand for all splines and selected to minimize the sum of square errors (SSE) for the central photodiode.

\begin{table*}
    \centering
    \begin{tabular}{llrr} 
        Model Name & Calibration Model Equation & Param. & Knots\\ \hline 
        Linear & $y = \theta_0 + \theta_1 x$ & 2 & 0\\ \hline
        Linear Spline & $y = \theta_0 + \theta_1 x + \theta_2 (x - k_1)_+$ & 3 & 1\\ \hline
        Linear 3 Spline & $y = \theta_0 + \theta_1 x + \theta_2 (x - k_1)_+\theta_3 (x - k_2)_+ + \theta_4 (x - k_3)_+$ & 5 & 3\\ \hline
        Quadratic Spline & $y = \theta_0 + \theta_1 x + \theta_2 x^2 + \theta_3 (x - k_1)_+^2$ & 4 & 1\\ \hline
        Quadratic 2 Spline & $y = \theta_0 + \theta_1 x + \theta_2 x^2 + \theta_3 (x - k_1)_+^2 + \theta_4 (x - k_2)_+^2$ & 5 & 2\\ \hline
        Cubic Spline & $y = \theta_0 + \theta_1 x + \theta_2 x^2 + \theta_3 x^3 + \theta_4 (x - k_1)_+^3$ & 5 & 1 \\ \hline
        Cubic 2 Spline & $y = \theta_0 + \theta_1 x + \theta_2 x^2 + \theta_3 x^3 + \theta_4 (x - k_1)_+^3 + \theta_5 (x - k_2)_+^3$ & 6 & 2 \\ \hline
    \end{tabular}
    \caption{Functional forms of the models considered for photodiode intensity calibration; $(x)_+$ is $\max(x, 0)$}
    \label{tab:models}
\end{table*}

Once an overall calibration model was selected, individual photodiode response curves needed to be fit. While OLS was used for initial model selection, its inability to simultaneously fit model parameters and knots made it a poor choice for final model tuning; handpicking knots for all 187 photodiodes was deemed impractical and would likely result in sub-optimal models. To overcome this shortcoming, final photodiode calibration functions were fit using MATLAB's \texttt{fminsearch} function, a numerical non-linear minimum finder that uses the simplex search method of Lagarias et al. \cite{lagarias1998convergence}. Both model parameters and knot positions were fit using \texttt{fminsearch} with one of three loss functions. The loss functions considered were the sum of square errors (SSE), the sum of square relative errors (SSRE), and the sum of absolute relative errors (SARE):
\begin{align} \label{eq:losses}
    \text{SSE} &= \sum_{i} \left[ f(x_i; \bm{\theta}) - y_i \right]^2, \\
    \text{SSRE} &= \sum_{i} \left[ \frac{f(x_i; \bm{\theta}) - y_i}{y_i} \right]^2, \\
    \text{SARE} &= \sum_{i} \left| \frac{f(x_i; \bm{\theta}) - y_i}{y_i} \right|,\end{align}
where the above sums are over recording, the $\{y_i\}$ are radiance values in physical units, the $\{s_i\}$ are the average measured photodiode intensities, and $f(s_i; \bm{\theta})$ is the selected model with parameter vector $\bm{\theta}$. A final fitting method was selected based on photodiode loss profiles and residuals plots. Finally, bootstrapped 95\% confidence intervals for each fitted model were computed to investigate model variance and bias. 1000 bootstraps were performed per model with each bootstrap creating a new 29 data point radiance-response vector by sampling the original data vector with replacement. Each of these 1000 data sets were individually fitted and its response curve was computed along the entire radiance range. Confidence intervals were then constructed by finding the 5$^{\text{th}}$ and 95$^{\text{th}}$ quantiles for the responses of the fitted models at each radiance value. 

\begin{figure}
    \centering
    \includegraphics[width=8.5cm]{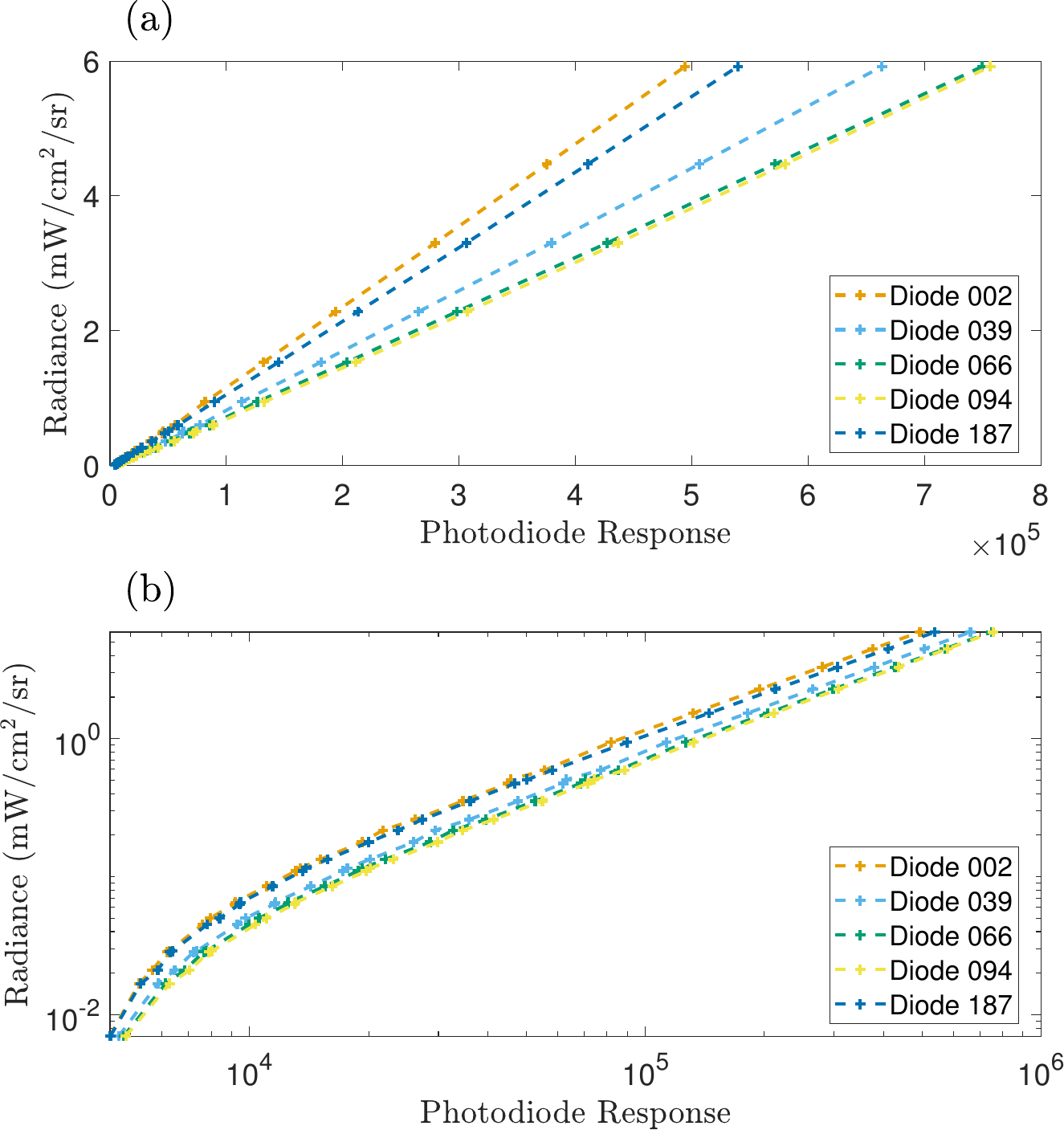}
    \caption{Photodiode to radiance response curves for five example photodiodes across the camera's hexagonal array. (a)Responses on a linear scale. (b) Responses on a log-log scale. Notice that the photodiode response is almost perfectly linear for high radiances, but takes on a notable nonlinear behavior for low radiances.}
    \label{fig:intensity_vs_radiance}
\end{figure}

We made measurements at 29 different radiances covering two orders of magnitude. The minimum measured radiance was $0\,{\rm mW}/{\rm cm}^2/{\rm sr}$, the maximum radiance was $5.92 \,{\rm mW}/{\rm cm}^2/{\rm sr}$, and the mean radiance was $0.751 \,{\rm mW}/{\rm cm}^2/{\rm sr}$. Figure \ref{fig:intensity_vs_radiance} shows the photodiode response versus radiance curves for a number of example photodiodes across the hexagonal array. It is clear that the diodes all share a similar response curve, with linear responses at high radiance values and non-linear responses for low radiances, but details differ, emphasizing the importance of calibrating each element independently.

\begin{figure}
    \centering
    \includegraphics[width=8.5cm]{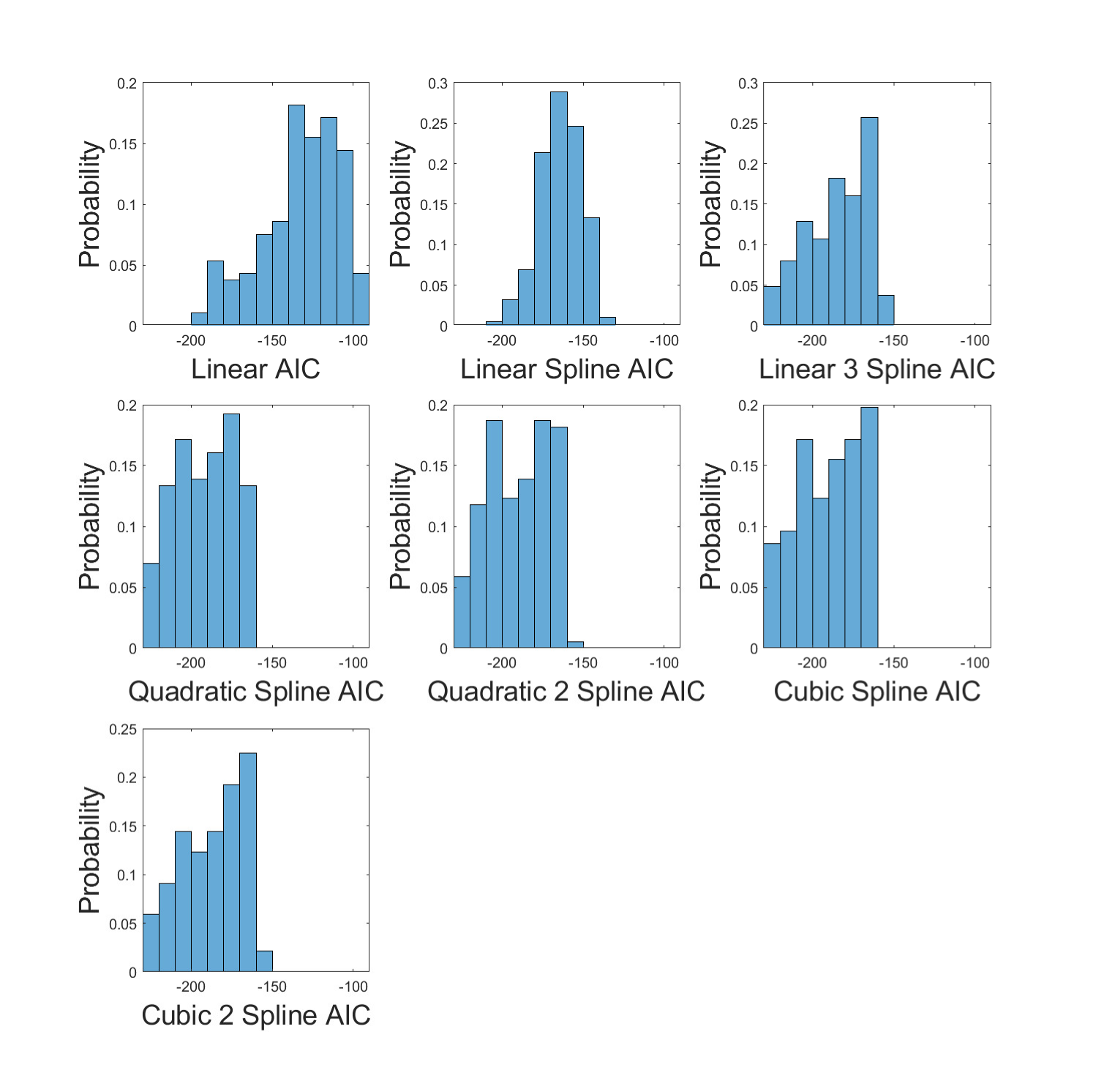}
    \caption{AIC distributions for each of the models listed in Table \ref{tab:models}. Note that the distribution changes dramatically across models types. The lowest mean AIC belongs to the Quadratic Spline model, while the Linear Spline has the lowest variance.}
    \label{fig:aics}
\end{figure}

\begin{figure}
    \centering
    \includegraphics[width=\linewidth]{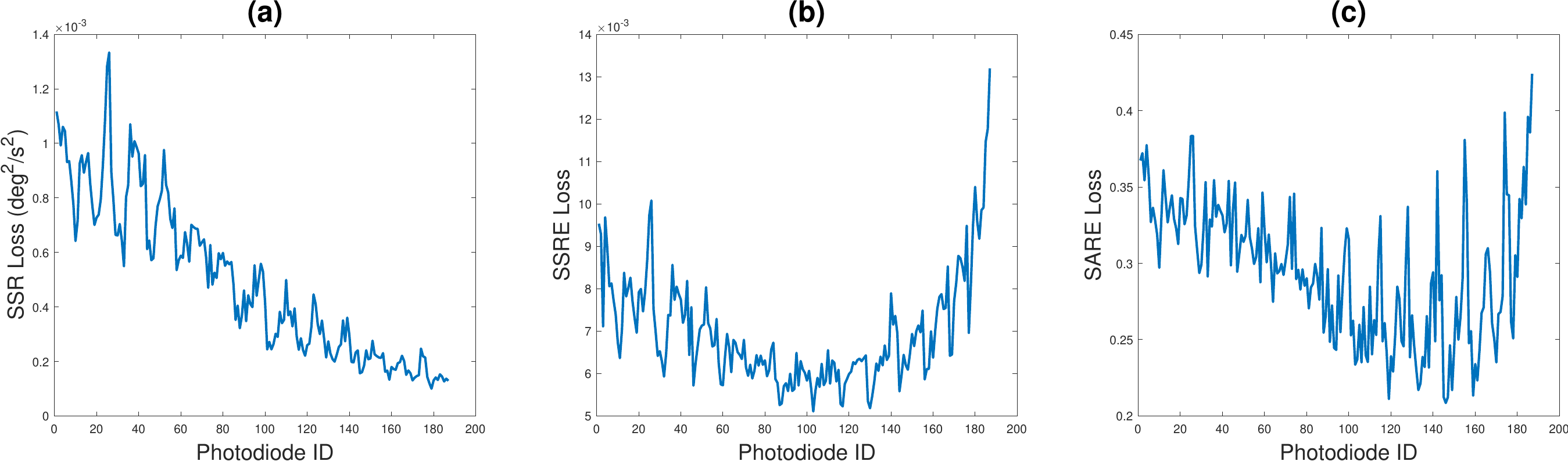}
    \caption{Quadratic spline losses across the SSR loss (a), SSRE loss (b), and SARE loss (c). Notice the SSRE loss displays the most radially symmetric profile.}
    \label{fig:spline_errors}
\end{figure}

Each of the models in Table \ref{tab:models} was fitted using OLS as described above and the Akaike information criterion (AIC) was computed for each fit. The AIC distributions across photodiodes are shown in Fig~\ref{fig:aics}. All models achieved a lower mean AIC than the baseline linear regression model ($\bar{x}_\text{AIC} = -133$). The one-knot quadratic spline (Quadratic Spline in Table \ref{tab:models}) had the lowest mean AIC ($\bar{x}_\text{AIC} = -192$) while the one-knot linear spline (Linear Spline in Table \ref{tab:models}) had the lowest AIC standard deviation ($s_\text{AIC} = 12.9$). The one-knot quadratic spline was selected as the overall calibration function given its lowest mean AIC and second lowest AIC variance. Every photodiode response curve was then fitted to three one-knot quadratic spline, one spline for each of the potential loss functions (Eq~\ref{eq:losses}). Figure \ref{fig:spline_errors} shows the final model losses for each metric across all diodes. Of particular note is that the SSRE loss function produces the most radially symmetric loss profile of the three functions (Fig~\ref{fig:spline_errors}b). 

\begin{figure}[b]
    \includegraphics[width=\linewidth]{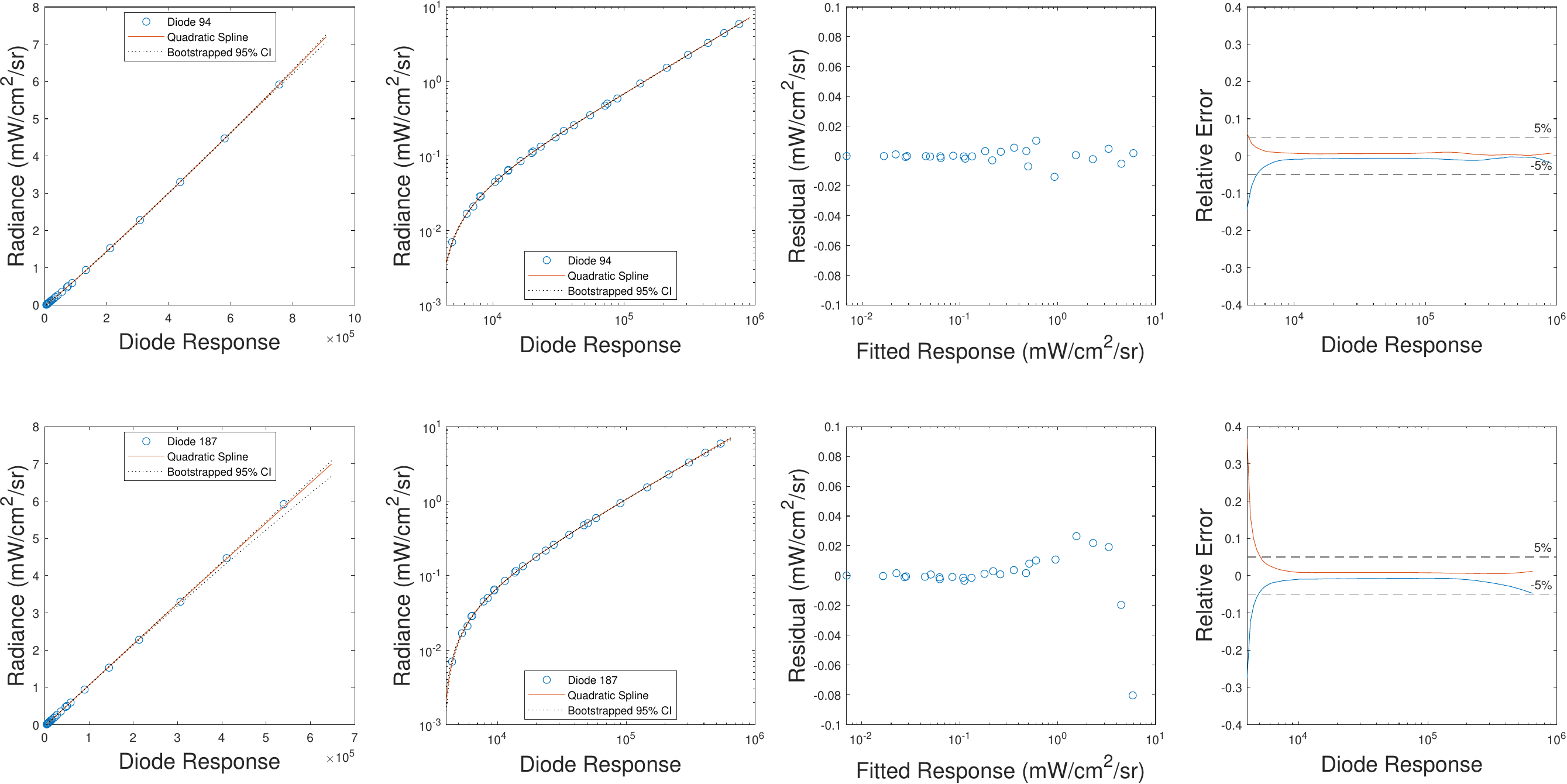}
    \caption{Fitted photodiode response curve for the central photodiode (top row) and edge most (bottom row) photodiodes, photodiode 94 and 187, respectively. The first column shows the fitted response curve on a linear scale while the second column shows the response on a log-log scale. The third column shows the residuals for each model while the last column shows the relative error of the bootstrapped 95\% confidence intervals across the entire response range. The fits for both photodiodes are extremely good, with residuals very close to zero. While a little non-linear trend remains for the edge most photodiode, the relative error of the bootstrapped 95\% confidence intervals for both photodiodes are below 5\% for all but the most extreme radiance values.}
    \label{fig:ssre_94}
\end{figure}

Further focusing on the SSRE models, Fig~\ref{fig:ssre_94} shows the fitted SSRE minimizing response curves for a photodiode in both the center and on the edge of the FlEye camera photodiode array. Photodiode 94's response curve, the central photodiode, is shown across the top row of Fig~\ref{fig:ssre_94}. The fitted model appears to closely follow the measured response curve for both low and high radiance regimes. The residuals are heteroskedastic but symmetric around zero. Additionally, the bootstrapped 95\% confidence intervals for the spline are both within 5\% relative error for all but the lowest measured radiances, light levels that are unlikely in natural scenes. Similarly, the response curve for an edge photodiode, photodiode 187, is shown across the bottom row of Fig~\ref{fig:ssre_94}. It again shows a good fit across response orders of magnitude. However, there appear to be some unfitted non-linear trend within the high radiance residuals. Even with this feature, the bootstrapped 95\% confidence intervals remain within the 5\% relative error range for all but the most extreme radiance values.

With the above results in mind, the one-knot quadratic spline was ultimately selected as the calibration function for its low average AIC, indicating it had the best average performance given its complexity. While the one-knot linear spline had a lower AIC variance, suggesting the model performed more consistently across photodiodes, the quadratic spline's improved performance was a more desirable feature. Additionally, quadratic splines have continuous first derivative at their knot points, a feature linear splines lack. This was an important consideration here as most visual estimators of velocity involve spatial and temporal gradients of the input visual signals. A final point in favor of the quadratic spline was the AIC distribution, which was both the most uniform of the higher complexity models and the lowest variance. Having ``outlier" photodiodes is extremely undesirable, especially since there is no clear reason why this would occur outside of an electrical or optical problem.  

Finally,  the symmetric errors of the SSRE fit models (Fig~\ref{fig:spline_errors}) suggest this loss function has the most consistent performance across photodiodes. As mentioned above, there is nothing that should break the radial symmetry of the FlEye camera's optics. Thus, it is surprising that the SSE minimizing model, and to a lesser extent the SARE minimizing model, produce such asymmetrical results. While the symmetry across photodiodes is not quantitatively measured here, it seems prudent to select the model with the most symmetrical results. Additionally, the focus on minimizing relative error insures models fit the non-linear response of the photodiodes at low radiance values as well as the high radiance linear response regime, where non-relative errors can be orders of magnitude larger than their low radiance counterparts. 

Another nice property of the SSRE fitted splines is the narrowness of their confidence intervals. Even for the photodiode with the largest loss value, photodiode 187, the 95\% confidence intervals of the fitted model were within the 5\% error range across all but the most extreme radiances (Fig~\ref{fig:ssre_94} bottom row). This suggests that the SSRE fitting process produces a relatively low variance estimator. While the SSRE minimizing model is definitionally not the least biased model (this honor belongs to the SSE minimizing model) the residuals for SSRE fitted models are small enough for the purposes outlined here. While some non-linear trend is still visible in the residuals of photodiode 187, the scale of these residuals is small enough that the trend can be effectively ignored. All of this taken together suggests that the SSRE minimizing one knot quadratic spline is a good model for calibrating the individual photodiode response profiles.

\vfill

\bibliography{fleye_camera_bibliography}

\end{document}